\documentclass[fleqn,usenatbib]{mnras}
\usepackage{bm}
\usepackage{newtxtext,newtxmath}

\usepackage[T1]{fontenc}
\usepackage[utf8]{inputenc}
\DeclareRobustCommand{\VAN}[3]{#2}
\let\VANthebibliography\thebibliography
\def\thebibliography{\DeclareRobustCommand{\VAN}[3]{##3}\VANthebibliography}

\newcommand{\youyou}[1]{#1}
\newcommand{\youyourev}[1]{#1}

\usepackage{graphicx}	
\usepackage{amsmath}	
\usepackage{subcaption} 
\usepackage{todonotes}
\usepackage{newunicodechar}
\newunicodechar{∼}{\textasciitilde}

\title[Geminga and Monogem in the CTAO Era]{Geminga and Monogem in the CTAO Era: Probing TeV Halos and Cosmic-Ray Transport}

\author[Li et al.]{
Youyou Li,$^{1}$\thanks{E-mail: y.li4@uva.nl}
Oscar Macias,$^{2}$
Manuela Vecchi$^{4}$
and Shin'ichiro Ando$^{1,3}$
\\
$^{1}$GRAPPA Institute, University of Amsterdam, 1098 XH Amsterdam, The Netherlands\\
$^{2}$Department of Physics and Astronomy, San Francisco State University, San Francisco, California 94132, USA\\
$^{3}$Kavli Institute for the Physics and Mathematics of the Universe, University of Tokyo, Chiba 277-8583, Japan\\
$^{4}$ Kapteyn Astronomical Institute, University of Groningen, Landleven 12, 9747 AD Groningen, the Netherlands
}

\date{Accepted XXX. Received YYY; in original form ZZZ}

\pubyear{\the\year{}}

\begin{document}
\label{firstpage}
\pagerange{\pageref{firstpage}--\pageref{lastpage}}
\maketitle
\begin{abstract}
TeV halos around Geminga (B0633+17) and Monogem (B0656+14) reveal strongly suppressed diffusion near middle-aged pulsars. Constraining the sizes and magnetic-field strengths of these slow-diffusion zones is essential for understanding lepton transport and the pulsars' contribution to the local positron excess. We construct two-zone diffusion models for both systems with \texttt{GALPROP v57}, in which electrons and positrons are injected according to the pulsar spin-down history and propagate through an evolving slow-diffusion bubble before entering the ambient interstellar medium. The models are calibrated to current HAWC and \textit{Fermi}-LAT measurements and forward-folded through the official Cherenkov Telescope Array Observatory (CTAO) instrument response functions to generate mock observations including diffuse and instrumental backgrounds. For 50-h exposures, both halos are detected at high significance, at approximately $13$--$30\,\sigma$, from either CTAO site. Under optimistic assumptions about the astrophysical background, CTAO can distinguish changes in the high-energy injection index of $\Delta\gamma_1 \simeq 0.2$ and magnetic-field differences of $\Delta B \simeq 2\,\mu\mathrm{G}$. For Monogem, compact slow-diffusion zones with $R_{\mathrm{SDZ}}\approx30\,\mathrm{pc}$ can be distinguished from configurations with $R_{\mathrm{SDZ}}\gtrsim50\,\mathrm{pc}$, whereas larger zones remain degenerate because of the finite field of view. Repeating the fits with mismatched Galactic diffuse templates changes the detection significances and parameter sensitivities by $\lesssim10\%$. CTAO observations will therefore provide stringent, though model-dependent, constraints on TeV-halo transport physics.
\end{abstract}

\begin{keywords}
gamma-ray -- cosmic-rays
\end{keywords}



\section{Introduction}
TeV halos are extended regions of TeV $\gamma$-ray emission surrounding middle-aged pulsars, typically reaching radii of $\sim$10–50 pc \citep{Linden2017}. While X-ray PWNe are generally confined to parsec scales \citep{Kargaltsev_2015, Izawa_2015}, TeV emission can extend much farther, making the observational distinction between evolved PWNe and particle halos non-trivial. Following \citet{Giacinti2020}, one may distinguish between PWN-dominated regions and extended halos based on whether the pulsar energetically dominates the local medium, rather than purely by spatial extent. The extended TeV $\gamma$-ray emission is generally interpreted as inverse-Compton (IC) radiation produced by multi-TeV electrons and positrons that escape the PWN and diffuse into the surrounding interstellar medium (ISM). Observations by HAWC have firmly established Geminga and Monogem as archetypal examples, revealing both their spectra and their remarkably extended morphologies \citep{2017_HAWC}. However, current $\gamma$-ray instruments are limited in complementary ways: water-Cherenkov detectors such as HAWC capture the full extent of these halos but lack the angular resolution needed to resolve their inner structure \citep{HAWC_2024}, while imaging air-Cherenkov telescopes can provide imaging down to arc-minute resolution, but face challenges when studying very extended emission due to their narrower fields of view and limited photon statistics, which often require integrating over angular scales larger than the nominal PSF, reducing the effective spatial resolution \citep{HESS_2023, P_hlhofer_2023,ALEKSIC2016}. As a result, the basic transport physics shaping these unusually large emission regions remains only partially constrained.

While diffusion is expected to be strongly suppressed within PWNe due to enhanced magnetic turbulence and particle confinement, recent observations suggest that this suppression may extend beyond the nebular boundary into the surrounding ISM. TeV halos around PWNe are therefore particularly intriguing, as their large spatial extent indicates that cosmic-ray (CR) transport remains inefficient over tens of parsecs outside the compact nebula. By analyzing PWN candidates from the 2HWC and HGPS catalogs, \citet{Di_Mauro_2020} inferred that the diffusion coefficient for $\sim 1$~TeV CRs in these regions is typically two orders of magnitude lower than the canonical Galactic average. Furthermore, \citet{Jóhannesson_2019} demonstrated that a two-zone diffusion model featuring a local slow-diffusion bubble surrounding the pulsar can successfully reproduce the observed TeV halo emission.
Since radiative energy losses are severe at these rigidities, such electrons and positrons naturally fail to propagate far into the ISM. For example, a 10~TeV electron has an energy-loss timescale of $\sim 50$~kyr in a $3~\mu\mathrm{G}$ magnetic field, \youyourev{comparable to the average Galactic magnetic field
strength} \citep{Beck_2011}. Under standard Galactic diffusion conditions, this corresponds to a propagation length of $\sim 670$~pc. This already indicates that any electron or positron observed at Earth at these energies is extremely sensitive to local injection and transport conditions. If the diffusion coefficient is suppressed by two orders of magnitude, as inferred in TeV halos, the corresponding diffusion length shrinks by roughly an order of magnitude. In that case, most high-energy leptons would lose nearly all of their energy near their sources, rather than contributing to the large-scale, diffuse Galactic CR population.

In the first LHAASO catalog of VHE and UHE $\gamma$-ray sources, 35 out of 90 sources pass their pulsar-association criteria, implying that a substantial fraction of catalog sources may be related to pulsars \citep{Cao_2024}. This means that pulsars constitute a significant population of TeV CR accelerators. Because part of the catalog may in fact be hadronic sources, the fractional contribution of pulsars to the TeV leptonic CR population is likely even higher. In addition, TeV halos appear to be a common feature around middle-aged pulsars \citep{Albert_2025,Hooper_2022}. Together, these findings make it urgent to understand the diffusion properties in the vicinity of such objects in order to obtain a complete picture of electron and positron transport in the Galactic plane.

Possible physical mechanisms responsible for the observed diffusion suppression include self-generated confinement driven by the propagation of electrons \citep{Malkov_2013,Evoli_2018,Schror_2022,Schroer_2025}, as well as environmental effects inherited from the parent supernova remnant and the larger wind-blown bubble or superbubble created by the progenitor system \citep{Nava_2016, Brahimi_2020}. \citet{Bourguinat_2026} argued that many middle-aged pulsars may remain within such progenitor-shaped environments for several hundred kiloyears. Theoretical studies predict that self-generated turbulence can efficiently suppress the diffusion coefficient within $\sim 20$~pc of the source, while environmentally driven turbulence may extend the suppression region to scales of up to $\sim 100$~pc. HAWC has demonstrated that a phenomenological model featuring a local slow-diffusion bubble of $\sim 25$~pc around the PWN can reproduce the TeV $\gamma$-ray emission observed around Geminga and Monogem \citep{2017_HAWC}. However, this represents only a lower limit based on the $\sim 10$~TeV $\gamma$-ray morphology. To robustly determine the size of the suppressed diffusion region, and to distinguish among physical mechanisms, observations of lower-energy inverse-Compton emission and corresponding synchrotron counterparts are essential.

The upcoming ground-based Gamma-ray telescope array: the Cherenkov Telescope Array Observatory (CTAO) will offer several key advantages for advancing TeV-halo studies, including arc-minute angular resolution at TeV energy, broad energy coverage from $\sim 20~\mathrm{GeV}$ to beyond $100~\mathrm{TeV}$, and significantly improved sensitivity over current instruments \citep{CTAbook2019}. \youyou{For sources such as Geminga and Monogem, TeV observations alone can be reproduced by multiple combinations of physical parameters, leading to degeneracies between quantities such as the size of the slow-diffusion region, the injected electron spectrum, and the local magnetic-field strength. Extending the measured energy range to lower energies can provide additional leverage on these parameters by probing the relative effects of particle diffusion and radiative energy losses. In this work, we investigate whether such measurements with the CTAO can help constrain these properties of TeV halos \citep{manconi_2024}.} 

In this work, we perform end-to-end simulations of CTAO observations to evaluate the detectability of TeV halos and the sensitivity to their underlying transport parameters, using Geminga and Monogem as benchmark sources. We adopt the two-zone diffusion framework in which electrons and positrons are injected by the parent PWN and propagate through a suppressed-diffusion region before entering the ambient ISM. The transport and associated IC emission are computed with \textsc{GALPROP} v57 \citep{Porter_2022}, \youyou{with the source injection power fitted} to reproduce current TeV-halo observations. We base the source modeling on the Monogem halo framework developed in \citet{Li_2025}, extend it here to the Geminga halo, and incorporate updated gamma-ray observation results from \citet{HAWC_2024}. We forward-fold these predictions through the CTAO instrument response functions and perform likelihood analyses to assess both detection significance and constraints on key model parameters such as injection spectrum, magnetic-field strength, and the size of the suppressed-diffusion region. The Gammapy package v1.2 is used in the CTAO simulation and data analysis \citep{gammapy:2023,gammapy:zenodo-1.2}.

The paper is structured as follows: in section \ref{sec: method}, we describe our modeling framework, including the source modeling, background modeling, and the forward-folding method for CTAO observation. In section \ref{sec: data analysis}, we then present results on CTAO detectability and robustness against background mismodeling, followed by forecasts for constraining key transport parameters.

\section{Methodology}\label{sec: method}
In this section, we provide details of the end-to-end modeling of the CR transport around Geminga and Monogem. We provide the gamma-ray emission prediction from GeV to TeV by utilizing the numerical code GALPROP v57. Furthermore, we simulate CTAO observations of the extended gamma-ray halo using the Gammapy package v1.2. From these mock data we assess the detection prospects and quantify the constraining power of the CTAO on the underlying transport properties, as described in Section~\ref{sec: data analysis}. 

\subsection{CR transport modeling Geminga and Monogem halos}
\label{sec:CR_transport} 
We consider a purely leptonic origin for the extended $\gamma$-ray emission. Electrons and positrons are accelerated in the PWNe—e.g., via diffusive shock acceleration or magnetic reconnection—and then propagate into the surrounding ISM. The dominant energy-loss processes for these leptons are synchrotron radiation and IC scattering. We assume electrons and positrons undergo the same acceleration, diffusion, and energy losses; therefore, hereafter, we model only the electron population. By charge symmetry, the injected pairs consist of equal numbers of electrons and positrons.

Because the acceleration power is ultimately limited by the pulsar’s spin-down power, we relate the lepton luminosity to the spin-down evolution:
\begin{equation}
    L_{e^{\pm}}(t)
    \;=\; \eta\,\dot{E}(t)\;=\; \eta\,\dot{E}_0\left(1+\frac{t}{\tau_0}\right)^{-\frac{n+1}{\,n-1\,}},
\end{equation}
where $\eta$ represents the fraction of the spin-down power at time $t$ that is injected into electrons and positrons, which we treat as a free parameter constrained by the observed surface brightness profiles of the halos. Here $\dot{E}_0$ is the initial spin-down power at $t=0$, and $\tau_0$ is the initial spin-down timescale. We adopt a braking index $n=3$ for both Geminga and Monogem, corresponding to the idealized magnetic-dipole model \citep{Lorimer_1997}.

The injected electrons and positrons follow a smoothed power-law spectrum \citep{Jóhannesson_2019}:
\begin{equation}\label{eq: injection}
    \frac{dn_{e^{\pm}}}{dE} \propto E^{-\gamma_{0}}
    \left[1+\left(\frac{E}{E_{\rm b}}\right)^{\frac{\gamma_{1}-\gamma_{0}}{s}}\right]^{-s}.
\end{equation}
At low energy the differential number density scales as $E^{-\gamma_0}$; at high energy it follows $E^{-\gamma_1}$, with a smooth transition at the break energy $E_{\rm b}$. We adopt a smoothness parameter $s=0.5$. Additionally, we fix the low-energy index to $\gamma_{0}=1.5$ and the break energy to $E_{\rm b}=100~\mathrm{GeV}$, which is typical of PWNe \citep{bucciantini_2010}. For the high-energy index $\gamma_1$, we explore \youyourev{$1.8,\,2.0,\,2.2,$ and $2.4$}. Note that Equation~\ref{eq: injection} does not include an intrinsic cutoff; we impose a high-energy cutoff by setting a highest electron/positron energy allowed in the simulation based on the HAWC $\gamma$-ray spectrum \citep{HAWC_2024} The photon cutoff at $\sim50$~TeV for the Geminga and Monogem TeV halos implies an electron cutoff at $E_{\max}\sim200$~TeV.

Lepton transport in the ISM is governed by diffusion and radiative losses; we assume no significant bulk advection in the Geminga/Monogem regions. The TeV $\gamma$-ray extension measured by HAWC constrains the mean distance electrons travel before losing most of their energy via IC scattering with the diffusion length given by
\begin{equation}\label{eq: diffusion length}
    R_{\rm diff}(E_{e})=\sqrt{4\,D(E_{e})\,\min\{\tau_{\rm cool},\,\tau_{\rm inj}\}},
\end{equation}
where $D(E_{e})$ is the diffusion coefficient at electron energy $E_{e}$. The shorter of
 the cooling time ($\tau_{\rm cool}$) and the injection time ($\tau_{\rm inj}$) sets the emission size. Using Eq.~\ref{eq: diffusion length}, we estimate the diffusion coefficient in the ISM around the PWNe, taking the injection timescale to be the pulsar age, $\tau_{\rm inj}=t_{\rm age}$. The initial spin-down timescale relates to the characteristic age, $\tau_{\rm c}$, as follows:
\begin{equation}\label{tau_0}
    \tau_{0}\equiv \frac{P_{0}}{(n-1)\dot{P}_{0}}
    = \frac{2\,\tau_{\rm c}}{(n-1)} - t_{\rm age}.
\end{equation}
Adopting $n=3$, and a typical $\tau_{0}=0.1\,\tau_{\rm c}$, yields
$t_{\rm age}=0.9\,\tau_{\rm c}$. Using $\tau_{\rm c}=342$~kyr (Geminga) and $110$~kyr (Monogem) \citep{Manchester_2005} gives
$\tau_{\rm inj}=307.8$~kyr and $99$~kyr, respectively.
Adopting the angular extensions measured by HAWC \citep{HAWC_2024}, the diffusion length of electrons producing 20~TeV $\gamma$ rays, corresponding to $E_e\sim100$~TeV,   is $R_{\rm diff}\simeq24.07$~pc (Geminga) and $24.18$~pc (Monogem). We note that these angular measurements were obtained from fitting an approximate surface-brightness profile; subsequent work has shown that the diffusion coefficients inferred from this approximation are consistent with those from more precise, morphology-aware methods \cite{HAWC_2024}. We compute $\tau_{\rm cool}$ from synchrotron and IC losses (see App.~\ref{appendix: cooling}). Synchrotron losses depend on the magnetic-field strength. Extended X-ray observations around Geminga and Monogem constrain $B$ in the halos. For Geminga, \cite{manconi_2024} finds $B\lesssim2~\mu\mathrm{G}$ from XMM-Newton flux limits and a NuSTAR non-detection. Degree-scale non-detections in eROSITA 0.5--2.0~keV data set upper limits of $B\lesssim1.4~\mu\mathrm{G}$ (Geminga) and $B\lesssim4~\mu\mathrm{G}$ (Monogem) \citep{Khokhriakova_2024}. To account for potential parameter degeneracies of the magnetic field with $D(E_{e},r)$, we explore magnetic-field strengths of $B=\{1,2,3\}\,\mu\mathrm{G}$ for both sources. 
Variation in $\tau_{\rm cool}$ with $B$ implies different $D(E_{e})$ via Eq.~\ref{eq: diffusion length}, to explain the $R_{\rm diff}$ implied from the HAWC observation. The corresponding $D_{100\,\rm TeV}$ values for different B assumptions are listed in Table~\ref{tab: diffusion_coefficient}.

\begin{table}
\centering
\begin{tabular}{lccc}
\hline
 & 1$\mu$G & 2$\mu$G & 3$\mu$G \\
\hline
$D_{100\ \mathrm{TeV}}$ (Geminga)   & 1.91 & 3.01 & 4.84 \\
$D_{100\ \mathrm{TeV}}$ (Monogem)   & 1.93 & 3.03 & 4.89 \\
$D_{\mathrm{PWN}}/D_{\mathrm{ISM}}$ (Geminga) & $1.23\times10^{-3}$ & $1.93\times10^{-3}$ & $3.11\times10^{-3}$ \\
$D_{\mathrm{PWN}}/D_{\mathrm{ISM}}$ (Monogem) & $1.24\times10^{-3}$ & $1.95\times10^{-3}$ & $3.14\times10^{-3}$ \\
\hline
\end{tabular}
\caption{Diffusion coefficient values at 100~TeV, expressed in units of $10^{27}\,\mathrm{cm}^2\,\mathrm{s}^{-1}$, and corresponding suppression ratios for Geminga and Monogem under different magnetic field strengths.}
\label{tab: diffusion_coefficient}
\end{table}

The energy scaling of the diffusion coefficient is: 
\begin{equation}
    D(E_{e})=D_{0}E_{e}^{\delta},
\end{equation}
where, $D_{0}$ is the normalization, and $\delta$ depends on the power spectrum of the plasma turbulence. We consider a Kolmogorov diffusion; therefore, $\delta$=0.35. The diffusion coefficients found around Geminga and Monogem are found to be 2--3 orders of magnitude suppressed compared to the Galactic average. The suppression rates obtained for different magnetic field assumptions are listed in Table \ref{tab: diffusion_coefficient}. To accommodate the different diffusion rates of CRs in the immediate ISM around the PWNe and the Galactic average, we adopted a two-zone diffusion model. The full energy and spatial dependence of the diffusion coefficient is then: 
\begin{equation}
D(E_e,r) = \left(\frac{E_e}{E_0}\right)^\delta
\begin{cases}
D_{\rm SDZ}, & r < R_{\rm SDZ}, \\
D_{\rm SDZ}\left( \dfrac{D_{\rm ISM}}{D_{\rm SDZ}} \right)^{\frac{r-R_{\rm SDZ}}{r_t-R_{\rm SDZ}}}, & R_{\rm SDZ} \le r \le r_t, \\
D_{\rm ISM}, & r > r_t .
\end{cases}
\label{eq:diffcoeff}
\end{equation}
Here, the region of the immediate ISM around the PWN with radius $R_{\rm{SDZ}}$, is called the `slow diffusion zone'(SDZ), where the diffusion coefficient is suppressed compared to the average ISM, referred to as $D_{\rm{SDZ}}$, as we calculated above. The radius $r_{t}$ is the transition radius, at which the diffusion coefficient smoothly transitions to the Galactic average value. To account for the gamma-ray emission extension, $R_{\rm{SDZ}}$ is at least equal to $R_{\rm diff}$, which is $\sim 25$~pc for both Geminga and Monogem. \youyou{A larger slow diffusion zone size is also consistent with observations at $\sim$10~TeV. In this case, the slower diffusion of lower-energy electrons over a larger radius around the PWN would produce more spatially extended IC and synchrotron emission, with enhanced surface brightness at larger radii, at photon energies below the TeV range.} Indeed, an analysis of the Geminga and Monogem halos with Fermi-LAT data places a lower limit of about 50 pc on the Geminga SDZ radius, while leaving the SDZ size around Monogem essentially unconstrained \citep{Di_Mauro_2019}. Theoretical investigations of self-generated bubbles by CR propagation indicate that the slow diffusion regions can extend up to $\sim100$~pc  \citep{Malkov_2013,Nava_2016,Evoli_2018,Schroer_2022}. Therefore, we explore the allowed space of the SDZ radius in steps of 20~pc. The values are found in Table \ref{tab:Pulsar_Model_parameters}. A dynamical model is considered for the evolution of the size of the SDZ. The $R_{\rm{SDZ}}$ values mentioned above are the radii of the SDZ today, while over the lifetime of the system, the bubble size grows as $R_{\rm SDZ}(t)=\mu\sqrt{t}$, where $\mu$ is a constant for each source. The transition radius $r_{t}$ follows the same evolution relation with a different value for the $\mu$ compared to that for $R_{\rm SDZ}$.
\begin{table*}
    \centering
    \small 
    \begin{tabular}{|p{4cm}|p{4cm}|p{4cm}|}
    \hline \hline 
    Parameter & Monogem Values & Geminga Values \\ \hline \hline
    Source & 2HWC J0700+143 & 2HWC J0635+180 \\
    Associated pulsar & B0656+14 (Monogem pulsar) & PSR J0633+1746 (Geminga pulsar) \\
    Initial spin-down power $(\dot{E}_0)$ & $1.84 \times 10^{35}$ erg s$^{-1}$ & $2.8 \times 10^{36}$ erg s$^{-1}$ \\
    Characteristic age $(\tau_{\rm c})$ & $110\;$kyr & $342\;$kyr \\
    Current age $(t_{\rm age})$ & 
    $99\;$kyr & $308\;$kyr \\
    Distance $(d)$ & $288\;$pc & $250\;$pc \\
     Proper motion $(\mu_{\alpha}\cos\delta,\mu_{\delta})$ &
    $(44.16,-2.43)~\mathrm{mas\,yr^{-1}}$ &
    $(107.5,142.1)~\mathrm{mas\,yr^{-1}}$ \\
    Galactic coordinates $(l, b)$ & $(201.1^{\circ}, 8.3^{\circ})$ & $(195.3^{\circ}, 3.8^{\circ})$ \\
    Galactocentric coordinates $(X, Y, Z)$) & $(-8.77, -0.10, 0.04)\;$kpc & $(-8.74 , -0.07 , 0.02)\;$kpc \\
    \hline
    Diffusion bubble core $(R_{\rm SDZ})$ & $30$, $\mathbf{50}$, $70$, $90\;$pc & $\mathbf{50}$, $70$, $90\;$pc \\
    SDZ transition radius $(r_{\rm t})$ & $50$,\textbf{70}, $90$, $110\;$pc & \textbf{70}, $90$, $110\;$pc \\
    Energy break $(E_{e,\mathrm{b}})$ &  $100\;$GeV & $100\;$GeV \\
    Energy max $(E_{e,\mathrm{max}})$ &  $200\;$TeV & $200\;$TeV \\
    Smoothness $(s)$ & $0.5$ & $0.5$ \\
    Low-energy spectral index $(\gamma_0)$ &  $1.5$ & $1.5$ \\
    High-energy spectral index $(\gamma_1)$ & $\mathbf{1.8}$,$2.0$, $2.2$, $2.4$ & $1.8$,$2.0$, $\mathbf{2.2}$, $2.4$ \\
    Magnetic field strength $(B)$ & $\mathbf{1}$, $2$, $3\;\mu$G & $\mathbf{1}$, $2$, $3\;\mu$G \\
    \hline
    \end{tabular}
    \caption{Parameters used to model the Monogem and Geminga halos and their slow diffusion zones. The parameter values in \textbf{bold} font are referred to as the benchmark parameter of each source.}
    \label{tab:Pulsar_Model_parameters}
\end{table*}

\subsection{Gamma-ray emission}
\subsubsection{Gamma-ray emission from Geminga and Monogem halos}
\begin{figure}
    \centering
    \begin{subfigure}[b]{0.5\textwidth}
        \centering
        \includegraphics[height=0.28\textheight]{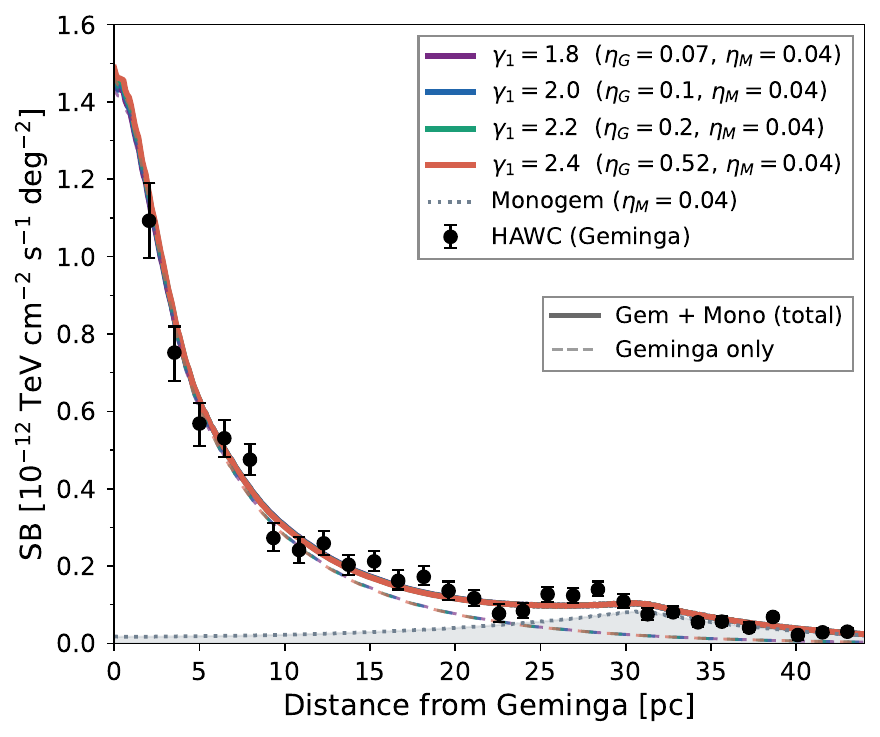}
     \end{subfigure}
     \hfill
     \begin{subfigure}[b]{0.5\textwidth}
         \centering
         \includegraphics[height=0.28\textheight]{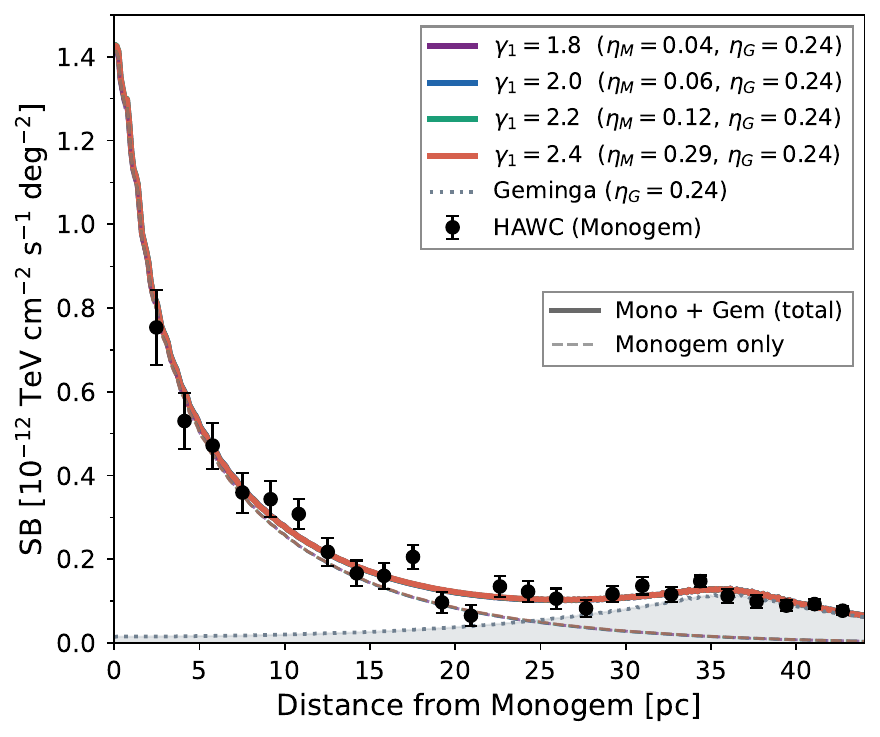}       
     \end{subfigure}
\caption{Surface brightness of IC emission in the $5.4$--$68.7$~TeV range for the 
Geminga halo (\textbf{top panel}) and the Monogem halo (\textbf{bottom panel}), each fitted jointly to account for the mutual contamination between the two sources. In both panels, curves correspond to different electron injection indices $\gamma_{1} = 1.8,\,2.0,\,2.2,\,2.4$ of the primary source, with magnetic field 
$B = 1~\mu$G and SDZ size $R_{\rm SDZ} = 50$~pc fixed. The injection efficiencies 
$\eta_{G}$ and $\eta_{M}$ of both sources are fitted simultaneously to the HAWC surface brightness data \citep{HAWC_2024}. The neighbour source (dotted) is held at a fixed benchmark model. The solid grey line shows the total combined surface brightness, and the dashed grey line shows the primary source contribution alone.}
     \label{fig: surface_brightness}
\end{figure}

We obtained the gamma-ray emission from the synchrotron radiation and IC scattering. The injection efficiency $\eta$ is obtained from fitting the surface brightness profile to the HAWC observation in the 5.4--68~TeV energy range. We show examples of the fitted profiles in Figure \ref{fig: surface_brightness}. For both Geminga and Monogem, assuming a magnetic field of 1$\mu$G, the required injection efficiency $\eta$ ranges from a few percent to a significant amount of the pulsar spin-down power. \youyou{A larger injection index} (softer injection spectrum) leads to a higher injection efficiency, consistent with previous modeling of the Monogem halo \citep{Li_2025}.

\begin{figure}
    \centering
    \begin{subfigure}[b]{0.48\textwidth}
        \centering
        \includegraphics[height=0.28\textheight]{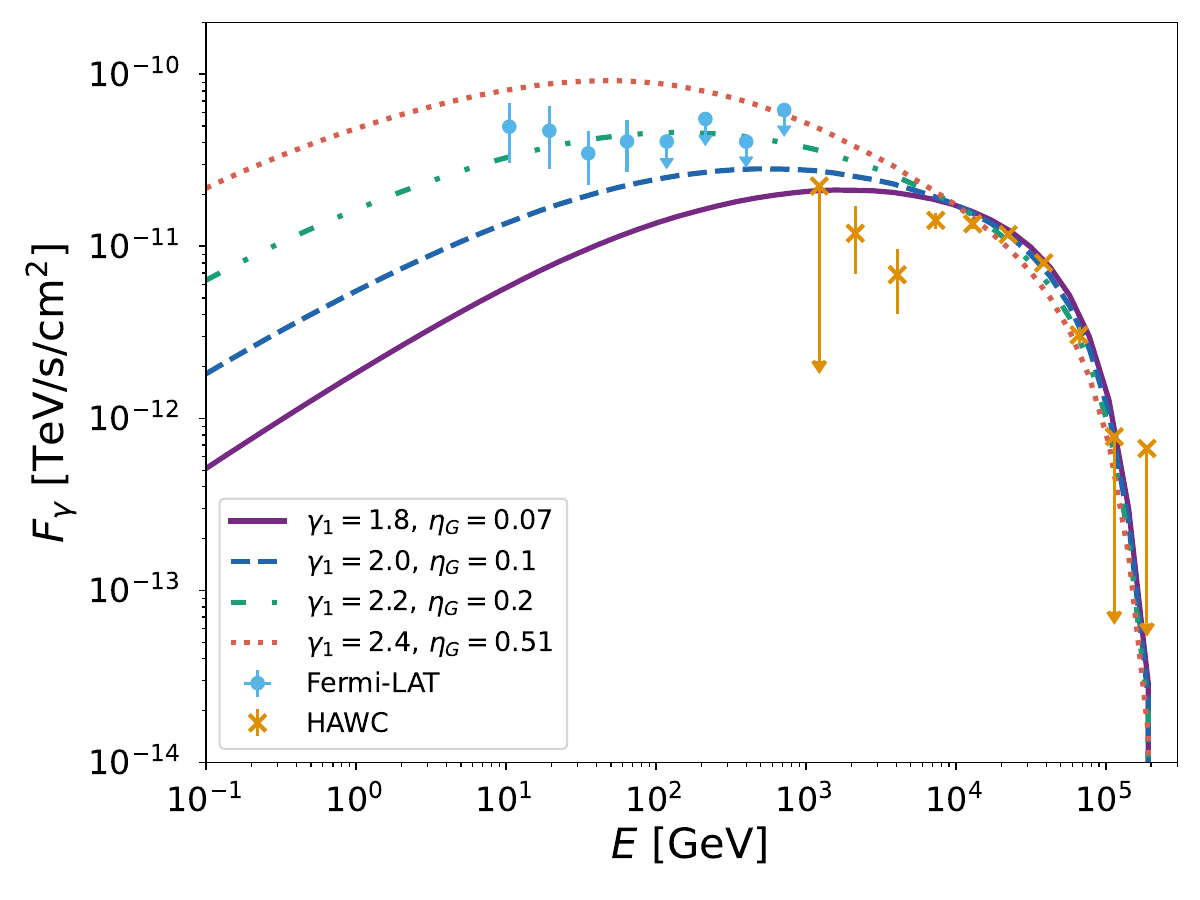}
     \end{subfigure}
     \hfill
     \begin{subfigure}[b]{0.48\textwidth}
         \centering
         \includegraphics[height=0.28\textheight]{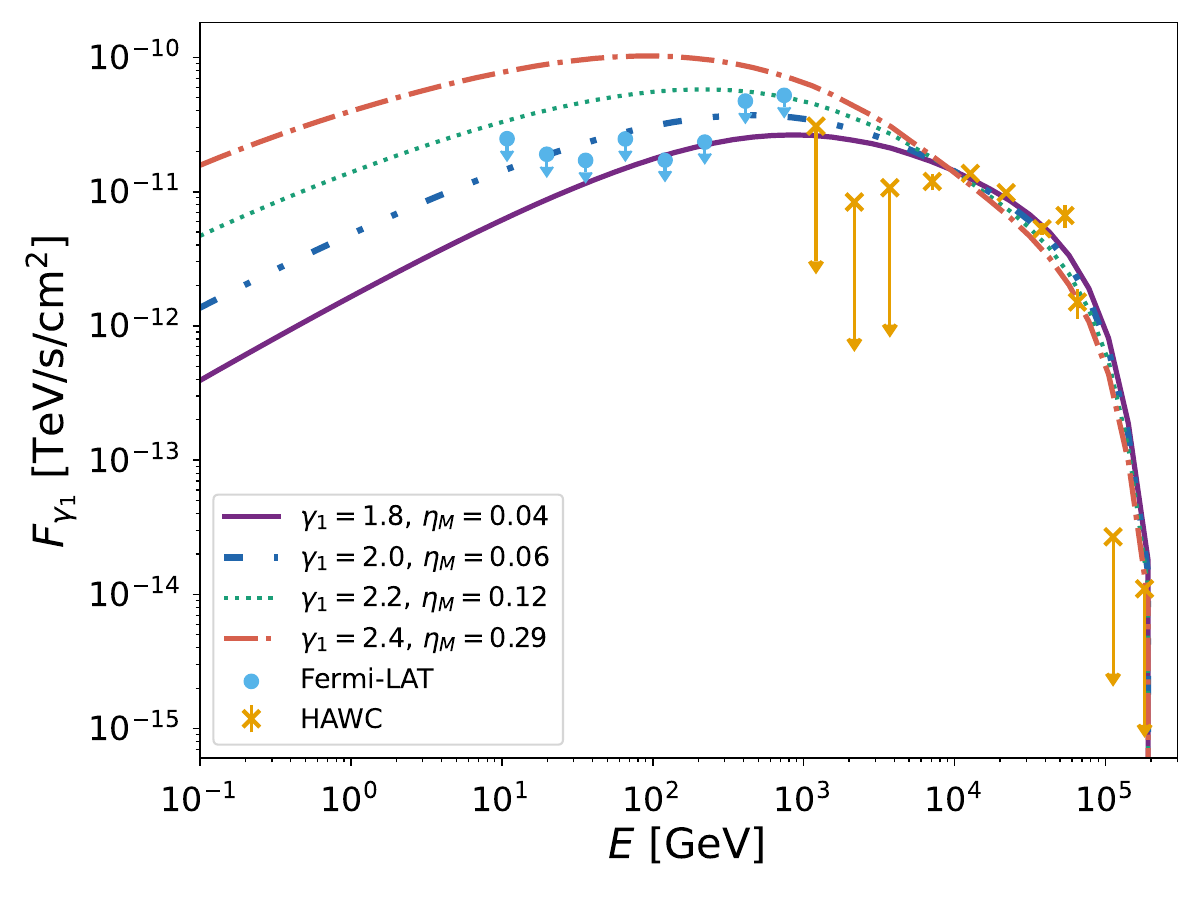}   
     \end{subfigure}
\caption{Energy spectrum averaged over 15$^{\circ}$ radius region around Geminga Halo (\textbf{top panel}), and Monogem Halo (\textbf{bottom panel}) for various electron injection index $\gamma_{1}$. We assume a benchmark value of magnetic field $B=1~\mu$G, and SDZ size $R_{\rm SDZ}=50$~pc. The spectrum is compared to Fermi-LAT data and upper limits \citep{Di_Mauro_2019}, and HAWC data \citep{HAWC_2024}. }
     \label{fig:ic_emission}
\end{figure}

We also compared our predicted energy spectrum with the observed data and upper limits to test the allowed range of source parameters. The IC spectra in the Fermi-LAT and HAWC energy ranges are shown in Figure \ref{fig:ic_emission}. We find that the predicted spectra are broadly consistent with the reported data points. We note, however, that both the Fermi-LAT and HAWC spectra are obtained through independent template-fitting analyses; thus, the derived spectra are model-dependent. The two studies adopt different source parameters—for example, the Fermi-LAT analysis obtained a best-fit injection index of 2.2–2.3, whereas the HAWC studies favor a much harder index of 1.0–1.1. Since our goal is to construct a different, but consistent, model that spans a wider energy range encompassing both the Fermi-LAT and HAWC regimes (and beyond), it is expected that our predicted spectra may not reproduce all data points simultaneously with perfect agreement. \youyou{As demonstrated in the H.E.S.S. data analysis of Geminga \citep{HESS_2023}, discrepancies between independently extracted spectra can be substantially reduced once differences in the assumed spatial morphology and extraction regions are properly accounted for.}   

\begin{figure*}
\centering
\begin{subfigure}[t]{.49\textwidth}
  \centering  \includegraphics[width=.8\linewidth]{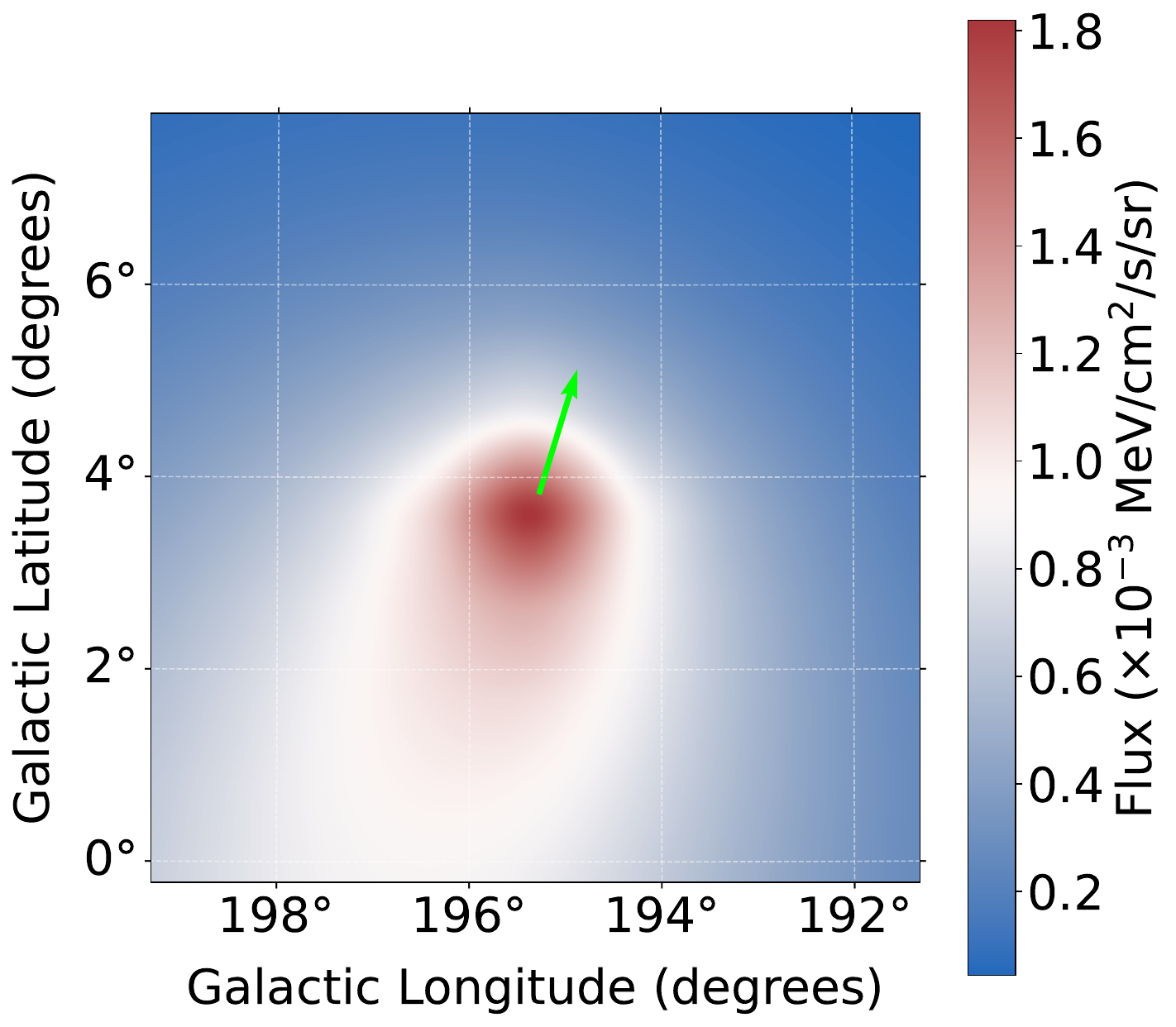}
  \caption{Geminga halo at 50~GeV}
  \label{fig:IC_pm_10GeV}
\end{subfigure}%
\begin{subfigure}[t]{.49\textwidth}
  \centering  \includegraphics[width=.8\linewidth]{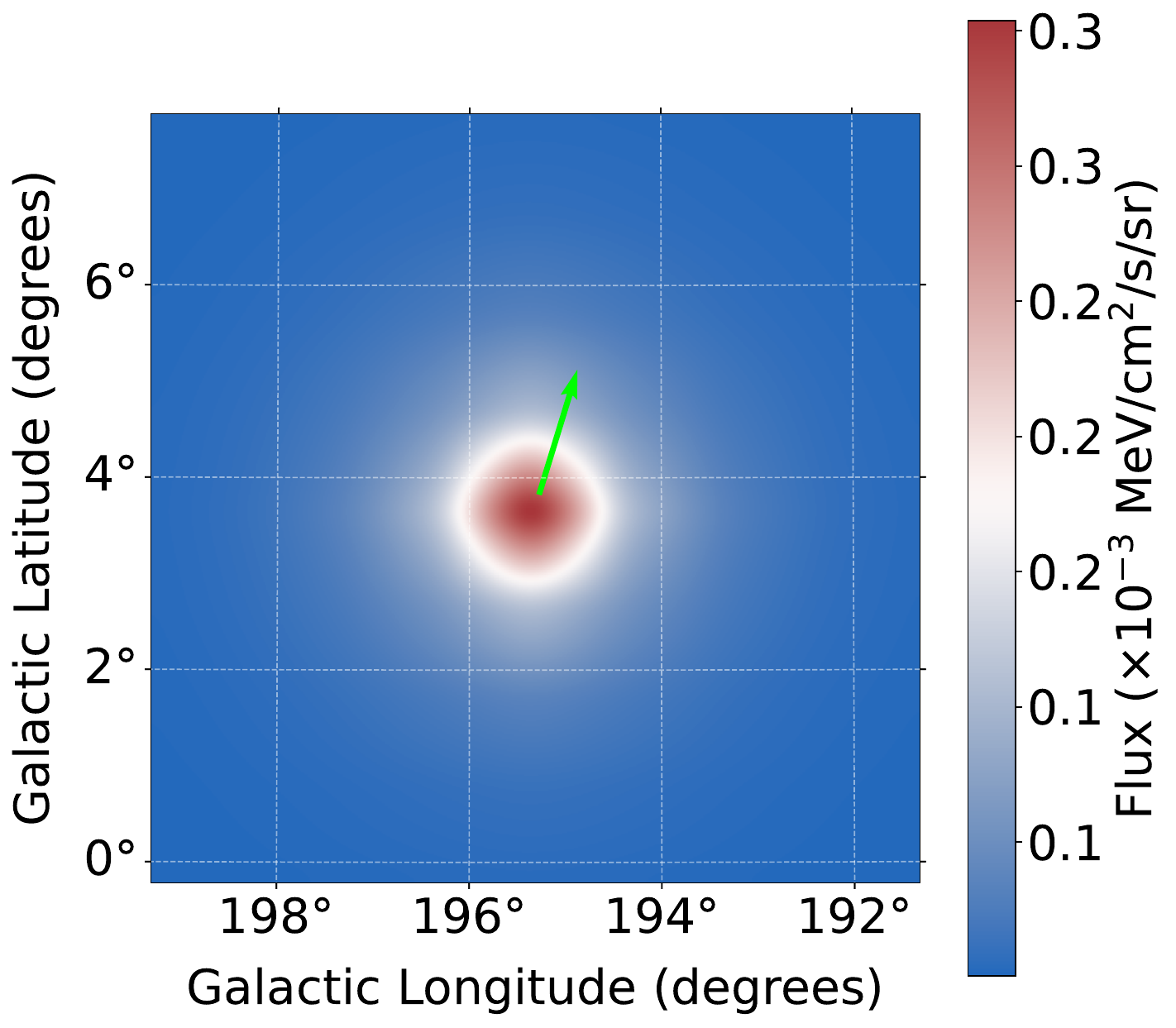}  \caption{Geminga halo at 100~TeV}
  \label{fig:IC_pm_10TeV}
\end{subfigure}
\begin{subfigure}{.49\textwidth}
  \centering  \includegraphics[width=.8\linewidth]{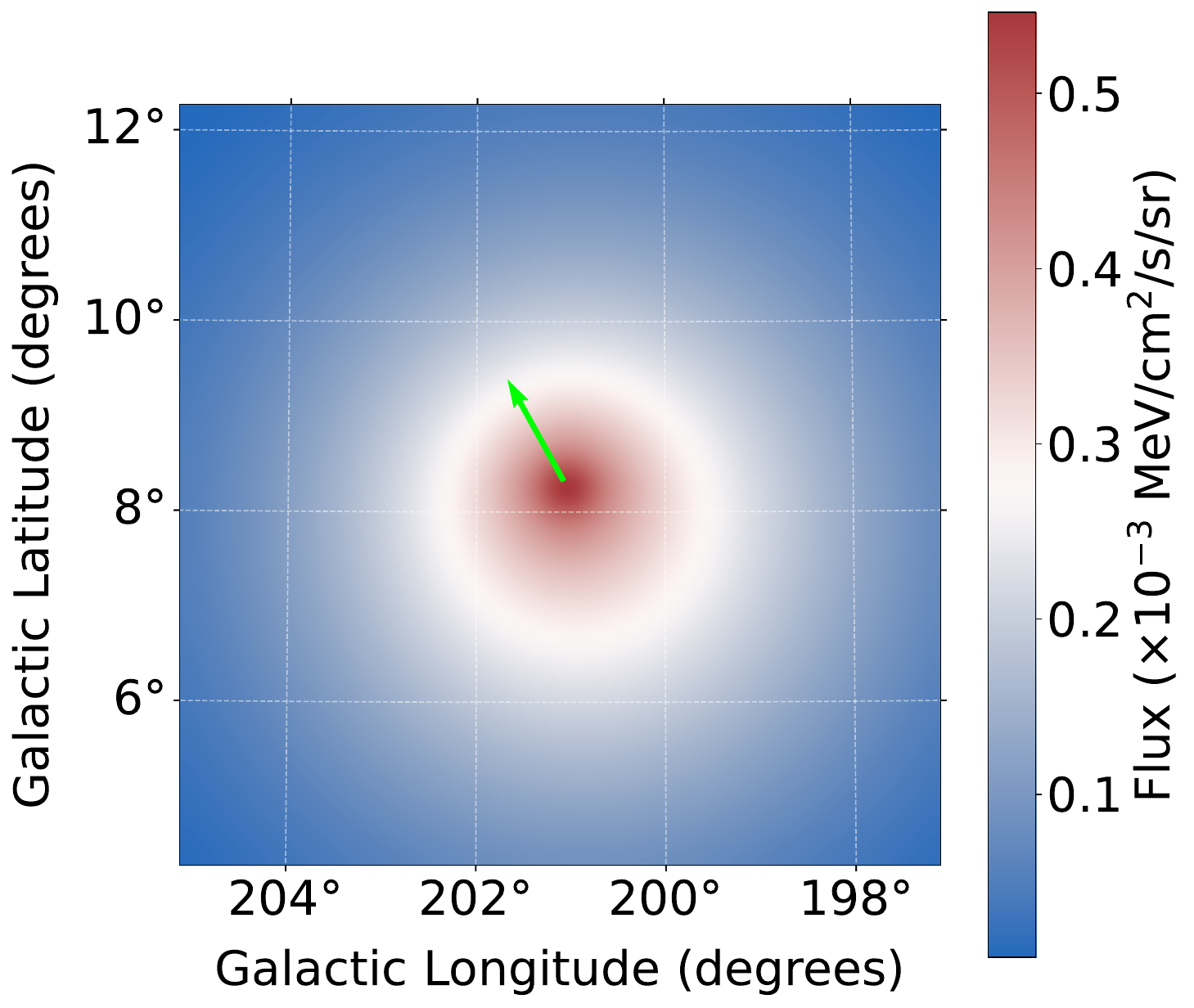}
  \caption{Monogem halo at 50~GeV}
  \label{fig:sync_pm_100GHz}
\end{subfigure}%
\begin{subfigure}{.49\textwidth}
  \centering  \includegraphics[width=.8\linewidth]{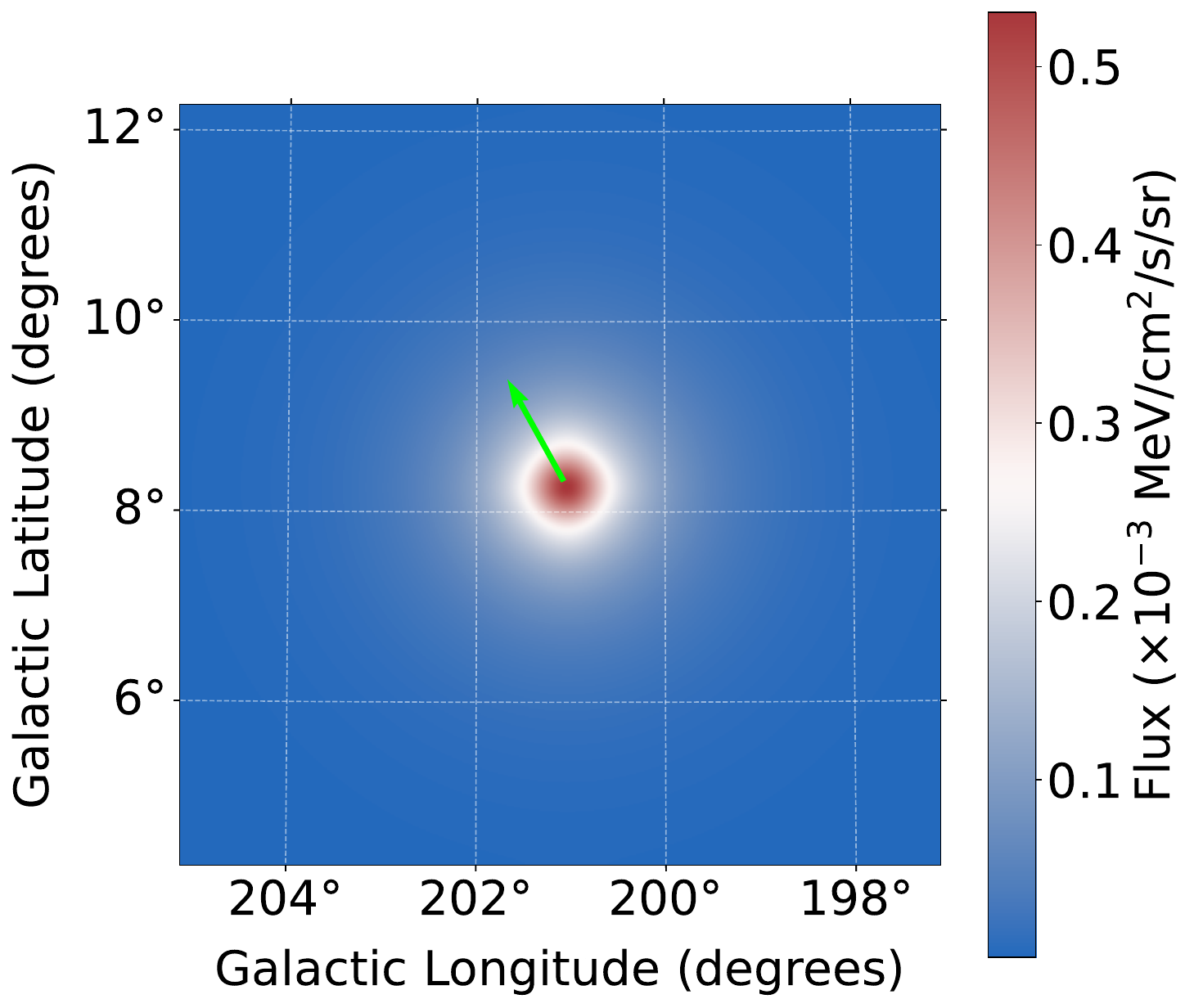}
  \caption{Monogem halo at 100~TeV}
  \label{fig:sync_emission_5keV}
\end{subfigure}\\
\caption{Intensity maps at 50~GeV and 100~TeV of Geminga (\textbf{top panels}) and Monogem (\textbf{bottom panels}) pulsars using benchmark parameter configurations. Geminga exhibits a large asymmetry in 50~GeV due to the pulsar's proper motion (panel (a)).  
}
\label{fig: intensity maps}
\end{figure*}

Our approach allowed us to model realistic energy and spatial emission dependence. In Figure \ref{fig: intensity maps}, we show the predicted surface-brightness morphology of the Geminga and Monogem halos at 50~GeV and 100~TeV, over an $8^{\circ}~\times~8^{\circ}$ FoV, comparable to the MST camera coverage. At 50 GeV, the Geminga halo exhibits a pronounced asymmetry caused by the pulsar’s high tangential velocity, whereas the much lower proper motion of Monogem leads to a more symmetric morphology. For both the Geminga and Monogem halo, the IC emission contracts as we increase the energy, meaning that the diffusion length of the electrons and positrons with energies $\gtrsim 1\,\rm TeV$ are limited by the energy loss time. The detailed energy-dependent morphology templates, which reflect the pulsar proper motion, and the diffusion property around the PWNe are crucial to identify the source properties in CTAO data.

\begin{figure}
    \centering
    \begin{subfigure}[b]{0.5\textwidth}
        \centering
        \includegraphics[height=0.28\textheight]{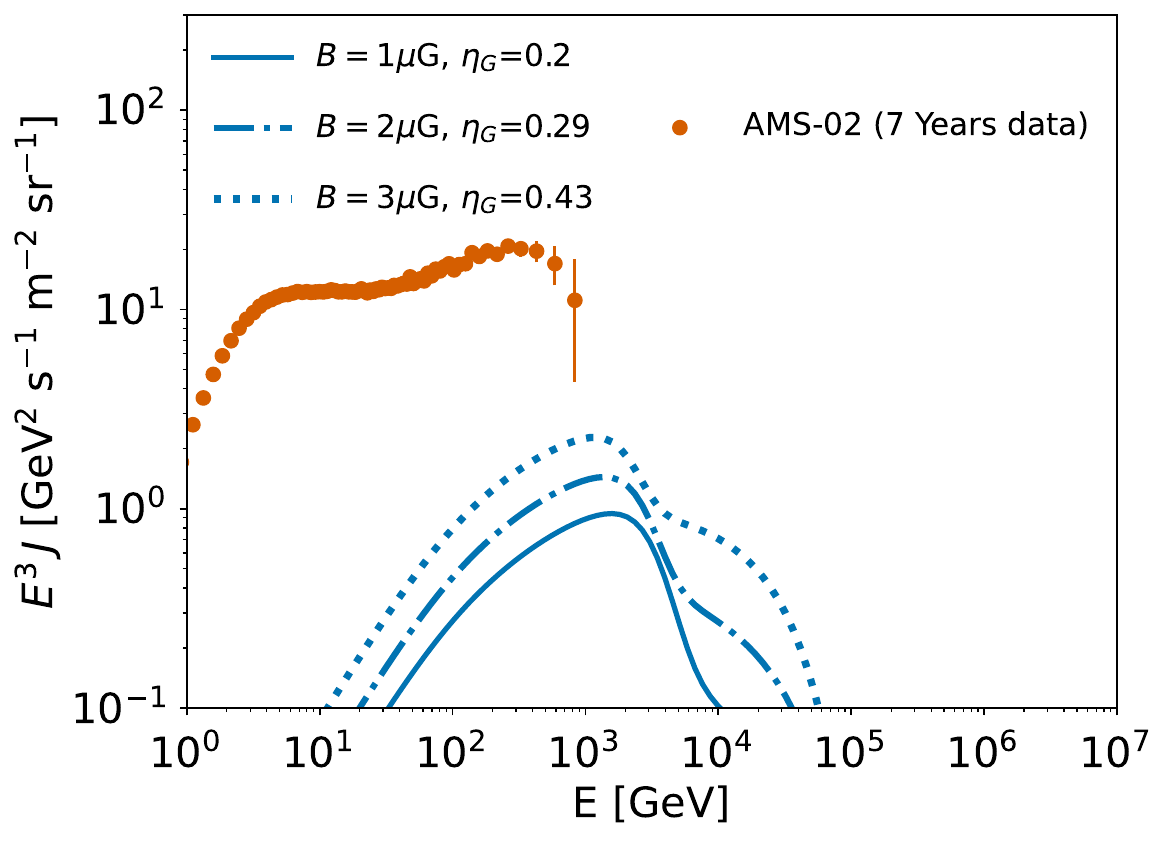}
     \end{subfigure}
     \hfill
     \begin{subfigure}[b]{0.5\textwidth}
         \centering
         \includegraphics[height=0.28\textheight]{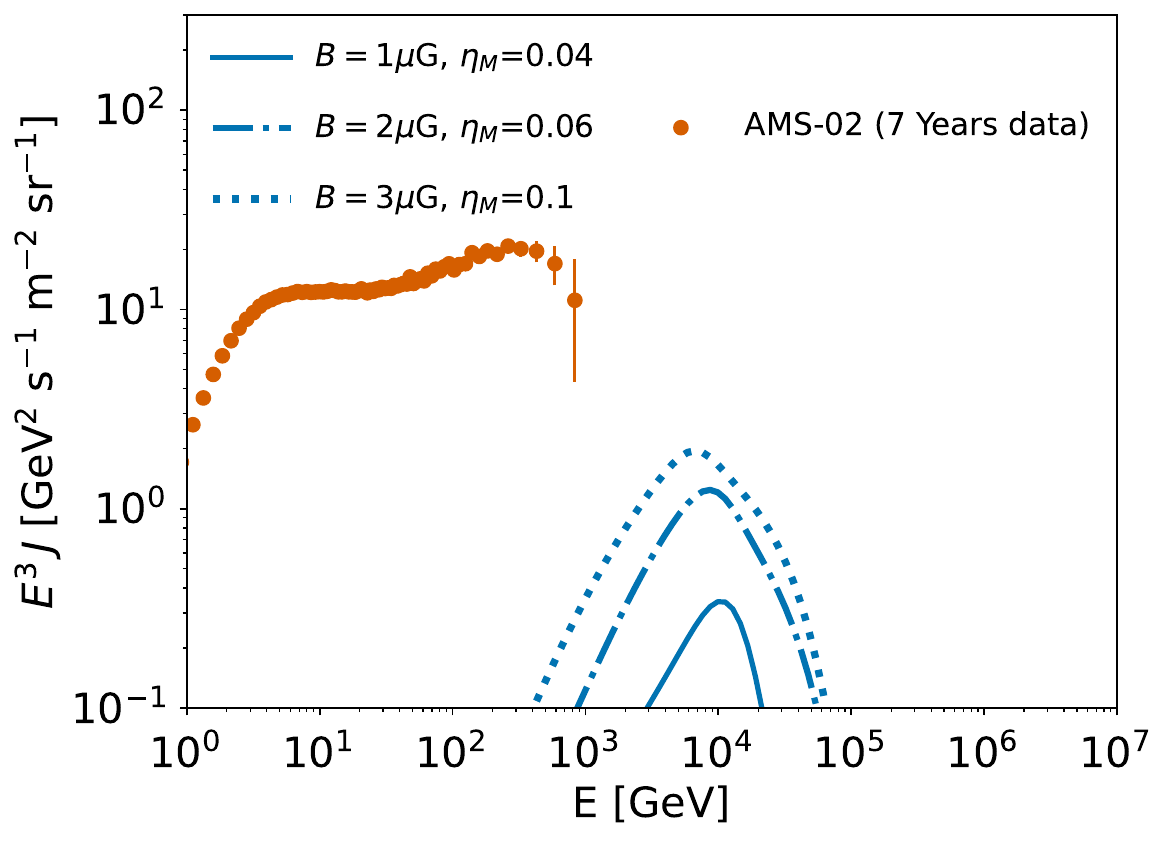}   
     \end{subfigure}
\caption{Positron flux expected at Earth from the Geminga halo (\textbf{top panel})
and the Monogem halo (\textbf{bottom panel}) for different magnetic-field strengths $B$. We assume a benchmark SDZ radius of $R_{\rm SDZ}=50$~pc and injection indices of $\gamma_{1}=2.2$ for Geminga and $\gamma_{1}=1.8$ for Monogem. The predicted positron flux is compared
to the 7-year AMS-02 measurement \citep{Aguilar_2013}.} 
\label{fig: positron_flux}
\end{figure}

Although the primary focus of this work is the $\gamma$-ray emission from TeV halos, nearby pulsars are also frequently discussed as plausible astrophysical candidates responsible for the positron excess observed above $100$~GeV \citep{Adriani_2008,Yuksel_2009,Dan_Hooper_2009}. For completeness, we therefore evaluate the expected positron flux at Earth from the Geminga and Monogem halos with our benchmark models. As shown in Figure \ref{fig: positron_flux}, for a slow-diffusion-zone radius of $R_{\rm SDZ}=50$~pc and injection indices of $\gamma_{1}=2.2$ for Geminga and $\gamma_{1}=1.8$ for Monogem, the contribution of these two sources individually accounts for up to $\sim 10\%$ of the measured positron flux above 100~GeV by AMS-02 after seven years of observation \citep{Aguilar_2013}. This level of contribution indicates that neither Geminga nor Monogem alone dominates the high-energy positron flux. 

\subsubsection{Gamma-ray background components}\label{subsec: gamma bkg}
To assess the detectability of the extended TeV halos with the CTAO, it is essential to model all relevant astrophysical and instrumental background components within the field of view (FoV). Since we adopt a forward-folded, template-based likelihood framework, the source emission and background contributions must be described consistently in both spatial and spectral dimensions. \youyourev{In this section}, we summarize the background components included in our model and their respective treatments.\\
We consider the following astrophysical and instrumental $\gamma$-ray background components:
(1) Galactic diffuse emission; (2) the isotropic diffuse background; (3) the residual (irreducible) CR background; (4) cross-contamination from Monogem in the Geminga FoV and \emph{vice versa}.

\begin{figure}
    \centering
    \begin{subfigure}[b]{0.5\textwidth}
        \centering
         \includegraphics[height=0.29\textheight]{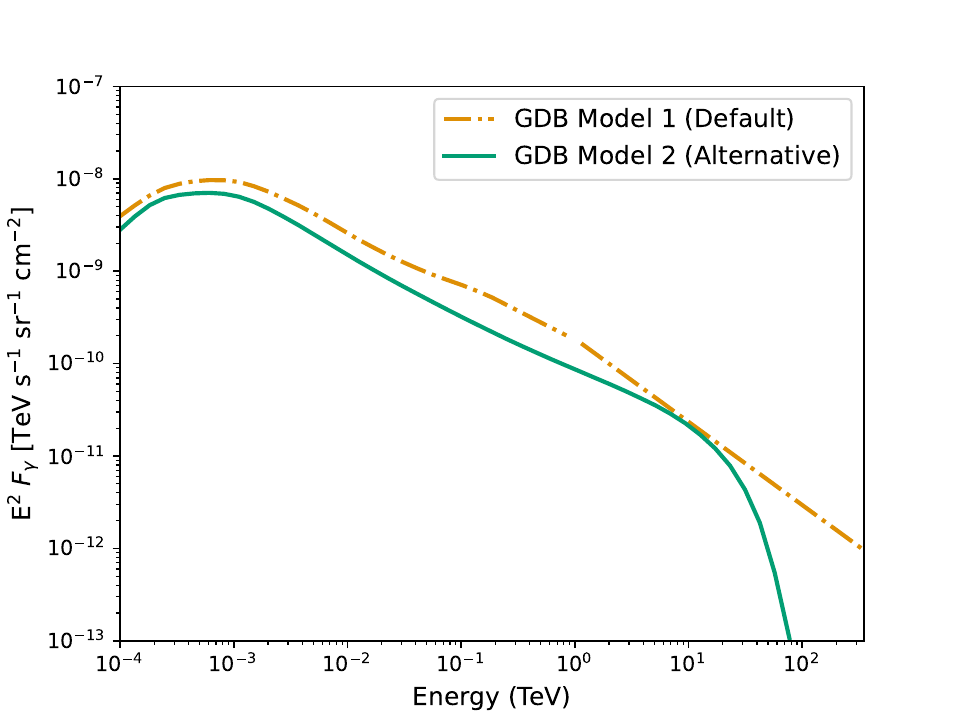}
     \end{subfigure}
\caption{Average diffuse $\gamma$-ray flux within a $2.15^{\circ}$ radius region centered on the Geminga halo, comparing the two Galactic diffuse background models used in this work. GDB Model~1 (default) follows the Fermi-LAT interstellar emission model with a TeV-scale extrapolation, while GDB Model~2 (alternative) adopts the 3D CR and ISRF-based diffuse emission model of \citet{Porter_2017}. The discrepancy between the two models increases toward multi-10~TeV energies.
}
\label{fig:GDB compare}
\end{figure}

\begin{figure}
    \centering
    \begin{subfigure}[b]{0.5\textwidth}
        \centering
         \includegraphics[height=0.29\textheight]{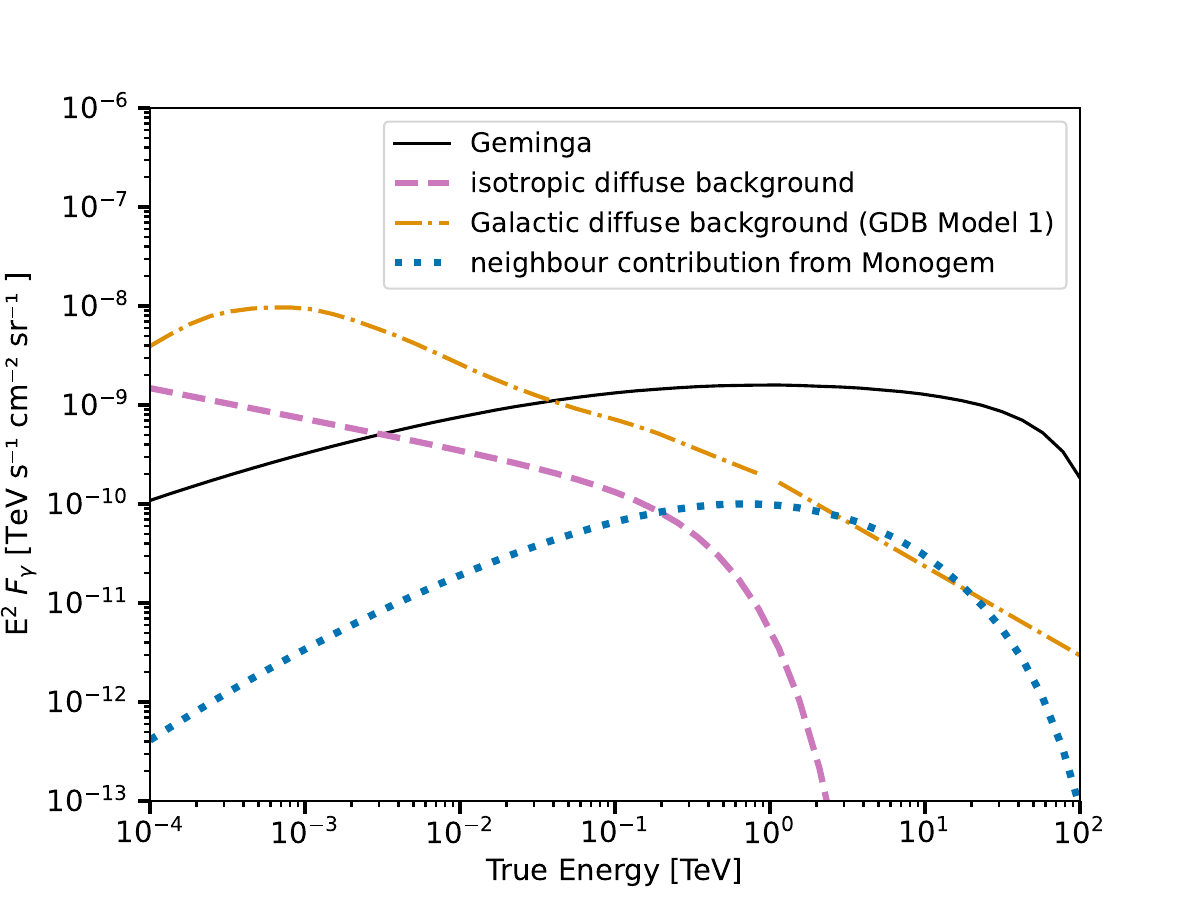}
     \end{subfigure}
     \hfill
     \begin{subfigure}[b]{0.50\textwidth}
         \centering
          \includegraphics[height=0.28\textheight]{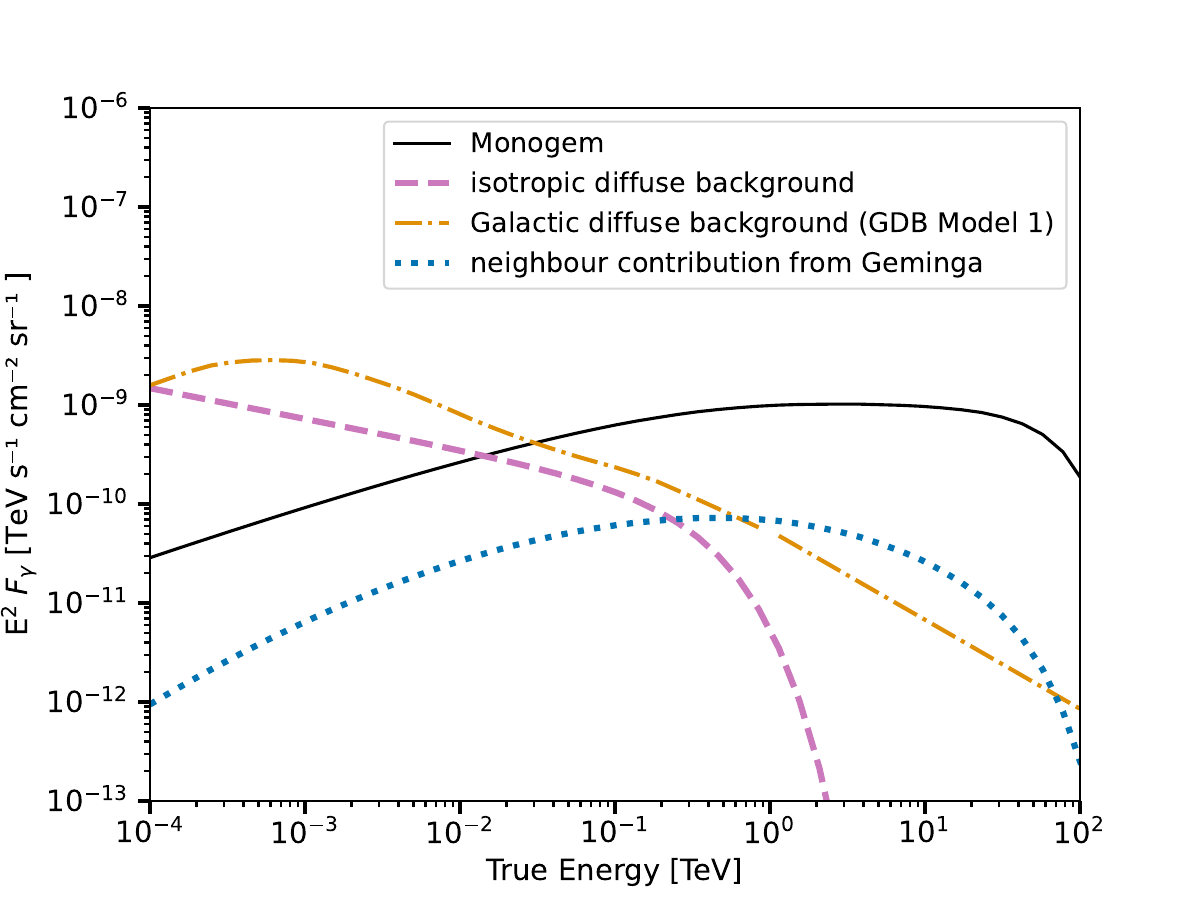}
     \end{subfigure}
\caption{Average flux of 2.15$^{\circ}$ radius region centered at the Geminga halo (top panel), and Monogem halo (bottom panel), compared to the intrinsic astrophysical background components. The source flux dominates the background components for both sources at above TeV energies.}
\label{fig:sourve_vs_bkg_flux}
\end{figure}

\begin{enumerate}
    \item \textbf{Galactic diffuse background (GDB).} 
    The properties of the Galactic diffuse $\gamma$-ray emission at TeV energies are uncertain, particularly away from the Galactic plane and in the vicinity of nearby CR sources. To assess the robustness of our results to background mismodeling, we therefore consider two different GDB prescriptions: a default model optimized to match current data, and an alternative physically motivated model with different spatial and spectral characteristics (see Section~\ref{section: background systematics}).
    \begin{itemize}
        \item \textbf{GDB Model 1 (default):} We adopt the Fermi-LAT Galactic interstellar emission model \texttt{gll\_iem\_v07.fits} up to $\sim 800$~GeV \citep{Abdollahi_2020}. Above this energy, we extrapolate the spectrum using a power-law index of $-2.5$, which softens to $-2.9$ above 1~TeV based on LHAASO measurements in the 1--25~TeV range \citep{Cao_2025}. The spatial morphology at TeV energies is assumed to follow that of the highest-energy LAT bin due to the absence of direct constraints.
    
        \item \textbf{GDB Model 2 (alternative):} We adopt the 3D diffuse $\gamma$-ray emission model of \citet{Porter_2017}, which self-consistently models CR densities, the interstellar gas distribution, and the 3D interstellar radiation field (ISRF). This results in a different morphology and spectral behavior at TeV energies compared to GDB Model~1, and therefore provides a useful test of the sensitivity of CTAO observations to realistic background uncertainties.
    \end{itemize}
    We compare the average diffuse $\gamma$-ray flux predicted by GDB Model~1 and GDB Model~2 within a $2.15^{\circ}$ radius region centered on Geminga in Figure~\ref{fig:GDB compare}. The two models show good agreement to a factor of 2--3 level up to $\sim 10$~TeV, while at higher energies the predictions begin to diverge significantly. This increasing discrepancy reflects the growing uncertainty in the Galactic diffuse emission at multi-10~TeV energies, where observational constraints are limited.
    
    \item \textbf{Isotropic diffuse background.} This component represents unresolved extragalactic emission. This component is not well measured beyond TeV energy. For consistency with the Galactic diffuse model, we adopt the analysis result based on Fermi-LAT data up to $\sim$800~GeV \citep{Abdollahi_2020}. A power-law spectrum of an index of 2.32, and an exponential cutoff at 279~GeV, normalized to an intensity of $I_{100~\rm MeV}=0.95\times10^{-7}$~MeV$^{-1}$~cm$^{-2}$~s$^{-1}$~sr$^{-1}$. As implied by its definition, this component has no spatial dependence.
    
    \item \textbf{Residual (irreducible) CR background.} These are non-$\gamma$ events that survive $\gamma$/CRs separation, dominated by electrons/positrons induced showers, with a small leakage of hadronic events. This background constitutes the irreducible instrumental background for IACTs. In Gammapy-based analyses, this is provided by the IRFs and folded through the exposure like other instrumental responses. For the CTAO we use the residual-background rates supplied with the IRFs (derived from detailed Monte Carlo simulations; see \citealt{BERNLOHR_2013}).\\
    \youyourev{To absorb residual mismodeling of the IRF-predicted instrumental background, we include a nuisance normalization parameter, $\alpha_{\rm irr}$, multiplying the residual CR background template in the likelihood. This term helps stabilize extended-source fits.}
    
    \item \textbf{Cross-source contamination (Geminga/Monogem).} When analyzing Geminga (Monogem), we include a template for Monogem (Geminga) within the FoV to account for flux leakage between the two extended halos and to avoid biasing the target source flux.

\end{enumerate}
To illustrate the relative contributions of the source and background components, we averaged the gamma-ray flux from the TeV halos and the background components \youyourev{over a 4.3$^{\circ}$ diameter} region around Geminga and Monogem, corresponding to approximately the LST FoV diameter. This choice is made for visualization purposes only and does not enter the likelihood analysis. The resulting intrinsic energy fluxes are shown in \youyourev{Figure ~\ref{fig:sourve_vs_bkg_flux}}.

\subsection{Simulated CTAO observations}\label{subsec: CTAO observatino}
\subsubsection{CTAO Configuration}\label{subsubsec: CTAO config}
We generate mock observations using the official CTAO \textit{Alpha Configuration}. In this layout, the northern array hosts 4 Large-Sized Telescopes (LSTs) and 9 Medium-Sized Telescopes (MSTs), while the southern array comprises 14 MSTs and 37 Small-Sized Telescopes (SSTs). The FoV diameters of the LST, MST, and SST cameras are approximately $4.3^{\circ}$, $7.5^{\circ}$, and $8.8^{\circ}$, respectively, corresponding to physical diameters of $\sim$19~pc, $\sim$33~pc, and $\sim$39~pc at the distance of Geminga ($d = 250$\,pc), and $\sim$21~pc, $\sim$40~pc, and $\sim$44~pc at the distance of Monogem ($d = 288$\,pc), \youyourev{consistently smaller than the explored diffusion bubble core sizes (see Table~\ref{tab:Pulsar_Model_parameters}), indicating that the full halo may extend beyond a single camera FoV, though a substantial fraction of the emission remains accessible within pointed MST and SST observations (see Figure~\ref{fig: surface_brightness})}. For each simulated data set, we adopt the publicly available instrument response functions \textbf{prod5 version v0.1} for the appropriate site, exposure time, and zenith angle \citep{CTAO_IRF_Prod5}.

To explore how site choice and exposure time affect the results, we test both the full Northern and Southern arrays with exposures of 5~h and 50~h. \youyou{The publicly available CTAO IRFs correspond to analyses optimized for specific observation durations: 0.5, 5, and 50~h. These IRFs include event-selection cuts tuned to maximize flux sensitivity for the corresponding exposure time. We therefore adopt the IRFs corresponding to the 5~h and 50~h observations in our simulations.} We note that the 5~h represents an extremely conservative exposure for a typical CTAO observation. It is included in the analysis to illustrate how performance scales with exposure. We simulate pointed observations of the Geminga and Monogem halos, placing each source at the center of the FoV. When observing from CTAO-North, the sources transit at zenith angles as low as $\sim 15^{\circ}$ in mid–January, while at CTAO-South, the minimum zenith angle is larger, $\sim 40^{\circ}$, reached in mid-March (see App. \ref{appendix: zenith}). Thus, observations from the Northern site allow access to significantly higher elevations for several months each year, providing improved sensitivity. The publicly released IRFs are provided for discrete zenith angles in steps of $20^{\circ}$ within the range $20^{\circ}$--$60^{\circ}$, we adopt the IRFs for the average zenith angle of the ranges: $40^{\circ}$ for observation at CTAO-North, and $60^{\circ}$ at CTAO-South.

\subsubsection{Mock data generation}\label{subsubsec: CTAO realizations}
We realize the CTAO observations of Geminga and Monogem pulsar by folding the sky model with the IRFs. IRFs depend on the true incoming gamma-ray energy, $E_{\mathrm{true}}$, and the offset angle from the detector center, $\theta_{\mathrm{true}}$. The detector responses include the effective area, $A_{\mathrm{eff}}(\theta_{\mathrm{true}},E_{\mathrm{true}})$, the point spread function, $\mathrm{PSF}(\theta \mid \theta_{\mathrm{true}}, E_{\mathrm{true}})$, the energy dispersion, $E_{\mathrm{disp}}(E \mid \theta_{\mathrm{true}},E_{\mathrm{true}})$, and the irreducible CR background, $\mathrm{Bkg}_{\rm irr}(\theta,E)$. Here, the irreducible CR background is treated as an instrumental response; thus, the expected irreducible gamma-ray counts distribution at reconstructed angle $\theta$ and energy $E$ in an observation of duration $t_{\mathrm{obs}}$ is: 
\begin{equation}\label{eq:bkg_count}
    N_{\mathrm{bkg}}(\theta, E)\,\mathrm{d}\theta\,\mathrm{d}E =\alpha_{\rm irr}  t_{\mathrm{obs}} \int_{E} \mathrm{d}E 
    \int_{\theta} \mathrm{d}\theta \;\mathrm{Bkg}(\theta, E).
\end{equation}
For astrophysical photons, the combined instrument response is given by:
\begin{align}
R(\theta, E \mid \theta_{\mathrm{true}}, E_{\mathrm{true}}) 
&= A_{\mathrm{eff}}(\theta_{\mathrm{true}}, E_{\mathrm{true}}) \nonumber \\
&\quad \times \mathrm{PSF}(\theta \mid \theta_{\mathrm{true}}, E_{\mathrm{true}}) \nonumber \\
&\quad \times E_{\mathrm{disp}}(E \mid \theta_{\mathrm{true}}, E_{\mathrm{true}}) .
\end{align}
\youyourev{The astrophysical sky model includes the target halo, Galactic diffuse emission, isotropic diffuse emission, and cross-source contamination from the neighboring halo. The residual CR background is treated separately as an instrumental component, following Eq. \ref{eq:bkg_count}.}
\begin{equation}
    \begin{aligned}
        \Phi_{\rm astro}
        \left(
            \theta_{\mathrm{true}},
            E_{\mathrm{true}}
        \right)
        &=
        \Phi_{\rm target}
        \left(
            \theta_{\mathrm{true}},
            E_{\mathrm{true}}
        \right)
        \\
        &\quad+
        \Phi_{\rm GDB}
        \left(
            \theta_{\mathrm{true}},
            E_{\mathrm{true}}
        \right)
        \\
        &\quad+
        \Phi_{\rm iso}
        \left(
            \theta_{\mathrm{true}},
            E_{\mathrm{true}}
        \right)
        \\
        &\quad+
        \Phi_{\rm contam}
        \left(
            \theta_{\mathrm{true}},
            E_{\mathrm{true}}
        \right).
    \end{aligned}
    \label{eq:astrophysical_sky_model}
\end{equation}

The expected gamma-ray photon counts in the detector, which are contributed from the incoming astrophysical flux, are obtained by convolving the flux from the sky region $\Phi_{\rm astro}(\theta_{\mathrm{true}}, E_{\mathrm{true}})$, with the response function:
\begin{align}
    N_{\mathrm{astro}}(\theta,E)\,\mathrm{d}\theta\,\mathrm{d}E 
    &= t_{\mathrm{obs}} \int_{E_{\mathrm{true}}} \mathrm{d}E_{\mathrm{true}} \int_{\theta_{\mathrm{true}}} \mathrm{d}\theta_{\mathrm{true}} \,
    R\!\left(\theta,E \,\middle|\, \theta_{\mathrm{true}}, E_{\mathrm{true}}\right) \nonumber \\
    &\quad \times \Phi_{\rm astro}(\theta_{\mathrm{true}}, E_{\mathrm{true}}) \,,
\end{align}
The total reconstructed counts are the sum of those contributed from the irreducible CR background and astrophysical flux:
\begin{equation}
N(\theta, E)\,\mathrm{d}\theta\,\mathrm{d}E 
= N_{\mathrm{bkg}}(\theta, E)\,\mathrm{d}\theta\,\mathrm{d}E 
   + N_{\mathrm{astro}}(\theta, E)\,\mathrm{d}\theta\,\mathrm{d}E,
\end{equation}

For every set of source parameters, we performed mock observations with exposure times of $t_{\mathrm{obs}} = 5$~h and $t_{\mathrm{obs}} = 50$~h at both CTAO-North and CTAO-South. For each such setup (i.e., a specific combination of source parameters, site, and exposure time), we generated 50 independent realizations using the CTAO IRFs. These realizations sample the intrinsic Poisson and background fluctuations expected for repeated observations and thus enable robust estimation of central values and uncertainties for all derived quantities.

\section{ANALYSIS AND RESULTS}\label{sec: data analysis}
\subsection{Likelihood analysis}\label{subsec: template fitting}
\youyou{We emphasize that our CTAO analysis adopts a forward-folded, template-based likelihood approach in which the source model and all background components are fitted simultaneously. The Galactic diffuse emission, isotropic background, residual CR background (provided through the instrument response functions), and cross-contamination from the neighboring halo are included as explicit components in the likelihood model. In this framework, no dedicated OFF region is required for background normalization within the FoV. This approach differs from traditional ON/OFF or reflected-region techniques commonly used in IACT analyses (e.g., in the H.E.S.S. detection of Geminga \citep{HESS_2023}), and is particularly well suited to extended sources whose spatial extent may cover a substantial fraction of the instrument FoV.}

For the analysis of extended gamma-ray sources observed by the CTAO, template fitting enables us to separate the morphological contributions of the source and background components, leading to more accurate inference of the source parameters \citep{Silverwood_2015}. We perform template fits for the Geminga and Monogem halos, respectively, to evaluate the detection significance and assess the distinguishability of different source parameters.

\youyourev{The model parameters are optimized by maximizing the likelihood, which corresponds to minimizing the Cash statistic. The total likelihood is evaluated as a product of Poisson probabilities over all spatial pixels and reconstructed energy bins:}
\begin{equation}
    \mathcal{L}=\prod_{p,k}{\rm Pois}(n_{pk}|\mu_{pk}),
\end{equation}
\youyourev{where $p$ and $k$ index the spatial pixels and reconstructed energy bins, respectively, $n_{pk}$ is the observed count, and $\mu_{pk}$ is the model-predicted count from the source and background templates.} To evaluate detection significance and model preference, we use the test statistic (TS) defined as:
\begin{equation}
TS = 2 \left( \ln \mathcal{L}_{\text{alt}} - \ln \mathcal{L}_{\text{null}} \right),
\end{equation}
where $\mathcal{L}_{\text{alt}}$ is the maximum likelihood under the alternative hypothesis (including the source model), and $\mathcal{L}_{\text{null}}$ is the likelihood under the null hypothesis (background only).\\

\subsection{Detection of the TeV halos}
Observing the Geminga and Monogem halos with CTAO-North and South each presents distinct advantages and challenges. The sources culminate at zenith angles of approximately $15^{\circ}$ at CTAO-North and about $40^{\circ}$ at CTAO-South. Observing from a large zenith angle leads to reduced sensitivity at low energies. However, CTAO-South includes the SSTs, which provide enhanced sensitivity above $\sim 5$~TeV---an energy range where the source flux begins to exceed the diffuse background. In addition, observations at larger zenith angles can increase the detection area for \youyourev{very-high-energy} showers. As a result, CTAO-South remains well-suited to probe the highest-energy component of the halo emission despite the low source altitude.

\begin{figure*}
    \centering
    \includegraphics[width=0.32\textwidth]{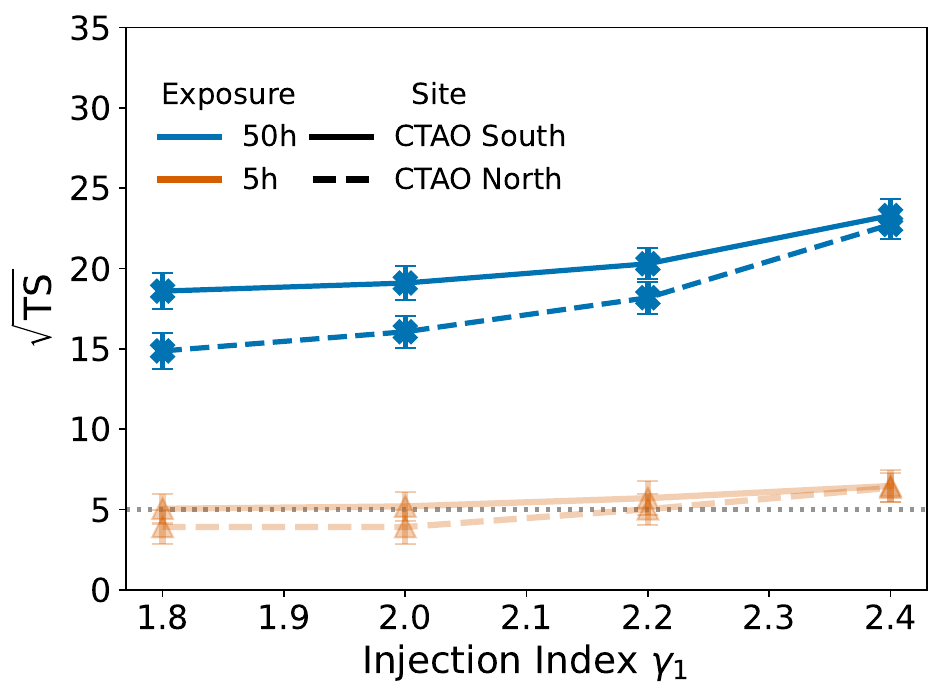}
    \includegraphics[width=0.32\textwidth]{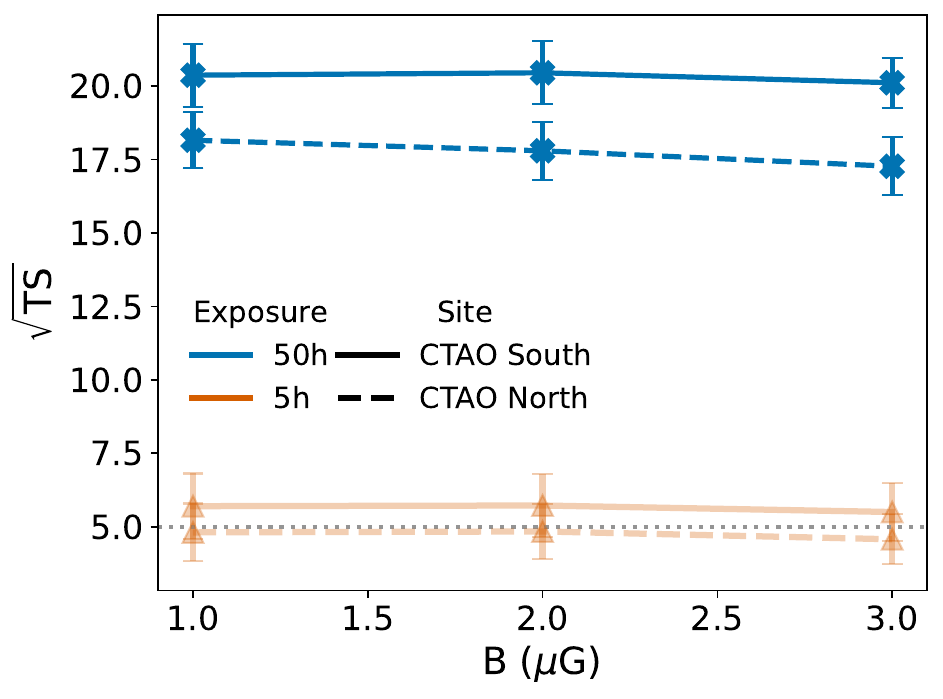}
    \includegraphics[width=0.32\textwidth]{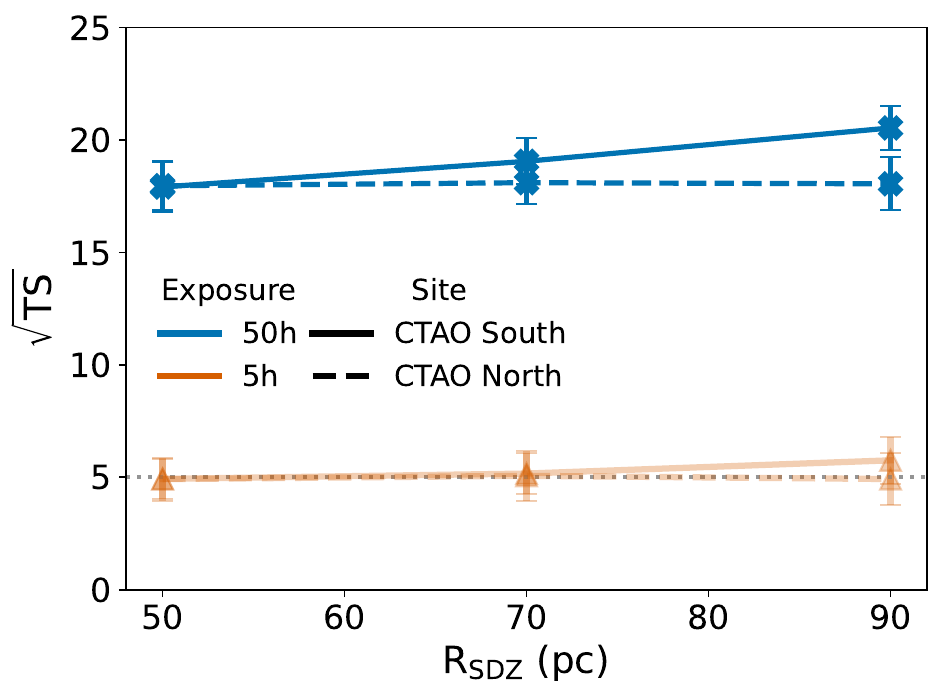}
    \caption{Detection significance of the Geminga halo by the CTAO. One model parameter is varied at a time while the remaining parameters are fixed at their benchmark values ($\gamma_1 = 2.2$, $B = 1\,\mu\mathrm{G}$, $R_{\rm SDZ} = 50$~pc). Results for a 5~h exposure are shown to illustrate how the detection significance improves with observing time; this represents an extremely conservative exposure for the CTAO. Each data point is derived from fitting 50 independent realizations of the same source model.}
    \label{fig:Geminga_detection_south}
\end{figure*}

\begin{figure*}
    \centering
    \includegraphics[width=0.32\textwidth]{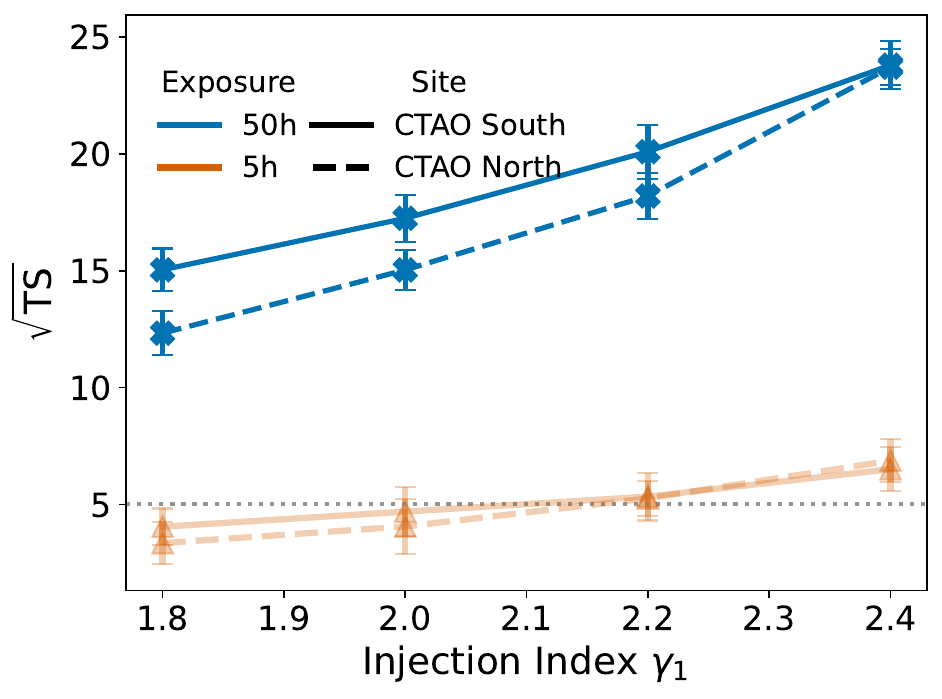}
    \includegraphics[width=0.32\textwidth]{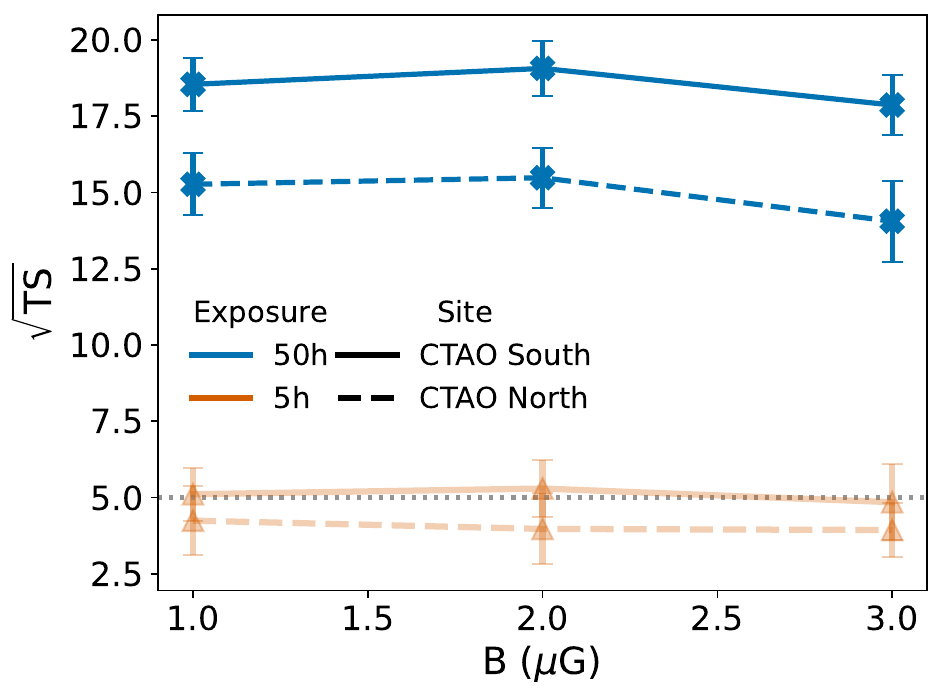}
    \includegraphics[width=0.32\textwidth]{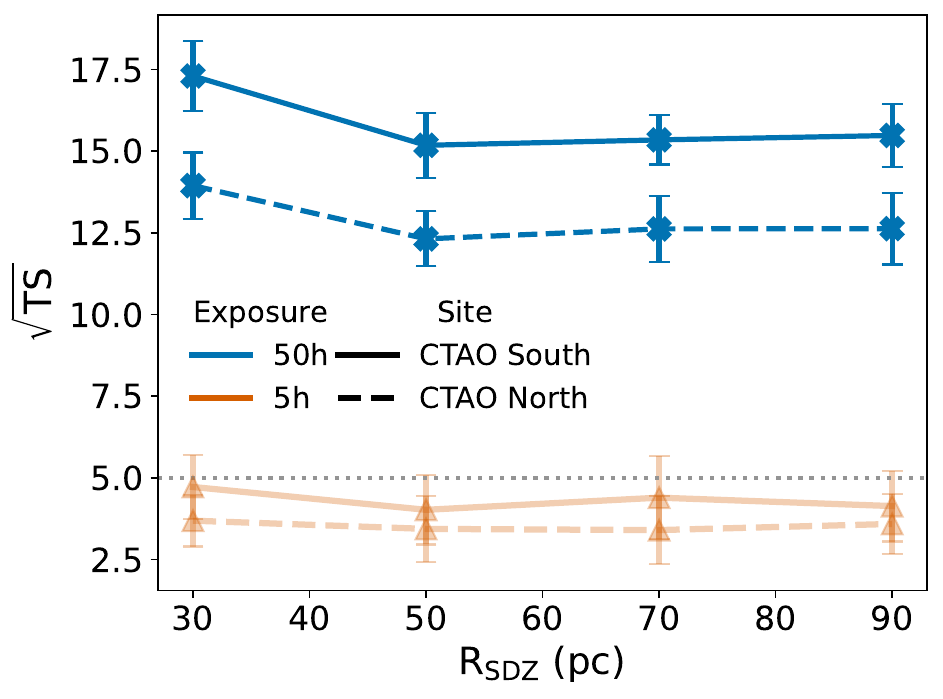}
    \caption{Detection significance of the Monogem halo by the CTAO. One model parameter is varied at a time while the remaining parameters are fixed at their benchmark values ($\gamma_1 = 1.8$, $B = 1\,\mu\mathrm{G}$, $R_{\rm SDZ} = 50$~pc). Results for a 5~h exposure are shown to illustrate how the detection significance improves with observing time; this represents an extremely conservative exposure for the CTAO. Each data point is derived from fitting 50 independent realizations of the same source model.}
    \label{fig:Monogem_detection_south}
\end{figure*}

In this analysis, the only additional free parameter in the alternative model is the normalization of the source flux, corresponding to one degree of freedom. According to Wilks' theorem, the detection significance is approximately given by $\sigma = \sqrt{TS}$. To assess the impact of individual source model parameters, we vary one parameter at a time while keeping the others fixed at their benchmark values. The detection results for the Geminga and Monogem halos are shown in Figure~\ref{fig:Geminga_detection_south} and Figure~\ref{fig:Monogem_detection_south}, respectively.

For both sources, a 50-hour exposure with the CTAO yields a detection significance of approximately $13$–$30\sigma$ across the explored ranges of injection indices, SDZ sizes, and magnetic-field strengths. The significance increases by roughly a factor of three when the exposure time is increased from 5~h to 50~h. As expected for background-dominated observations, the detection significance approximately scales with the square root of the exposure time ($\propto \sqrt{t}$). The two exposure times considered here represent short and deep observations, and the results can be extrapolated to other exposure times using the same scaling.

CTAO-South generally provides higher detection significance than CTAO-North for both Geminga and Monogem. \youyou{However, the two sites probe complementary energy regimes due to both the observing conditions and the array configuration. CTAO-North observes the sources at smaller zenith angles and benefits from the LSTs, which extend the detection capability to lower energies and allow a closer connection with the Fermi-LAT energy range. At these energies, the halo emission is expected to be very extended, potentially exceeding the FoV of the LSTs, which may complicate the characterization of the full spatial morphology and the imprint of pulsar proper motion. In contrast, CTAO-South includes the SSTs, which extend the detection capability to multi-TeV energies together with a larger FoV that is well suited for studying extended emission. However, at these higher energies, the halo emission becomes more concentrated around the source, reducing the observable impact of pulsar proper motion. Characterizing the very extended low-energy halo with CTAO-North, particularly to capture the effects of pulsar proper motion, may therefore require dedicated analysis strategies such as \youyourev{grid pointings} (see Section \ref{section: background systematics}).}

A positive correlation between detection significance and the injection index is also observed for both sources. This occurs because a softer injection spectrum (higher $\gamma_1$) requires a larger electron injection efficiency $\eta$ to reproduce the observed $\gamma$-ray flux. The resulting increase in injected power enhances the IC emission in the 10~GeV–1~TeV band (see Figure~\ref{fig:ic_emission}), thereby improving the statistical significance of the CTAO detection. With an injection index of $\gamma_1 = 2.4$, the suppressed high-energy flux and enhanced low-energy flux reduce the relative advantage of the CTAO-South so that CTAO-North and CTAO-South yield similar detection significance.

We find little to no dependence of the TS on the magnetic field strength within the range of $1$--$3~\mu$G, and on the diffusion zone size beyond 50~pc for Geminga and 30~pc for Monogem. In particular, the magnetic field does not significantly alter the spectrum above GeV energies. Although a stronger magnetic field requires a higher injection efficiency to maintain the TeV gamma-ray flux normalization, it also leads to increased synchrotron losses, resulting in a higher synchrotron flux, while this feature is outside of the energy range of the CTAO.

\subsubsection{Robustness against background uncertainties}\label{section: background systematics}
\textbf{Galactic diffuse background}\\
The results above are obtained by fitting the simulated observations with the same sky models used to generate the data, implicitly assuming perfect knowledge of both the source and the diffuse background. \youyou{In real CTAO analysis, this assumption is unlikely to hold, particularly due to uncertainties in the Galactic diffuse emission and the irreducible CR background. We therefore assess the robustness of the detection significance against diffuse-background mismodeling, and discuss strategies for calibrating the irreducible cosmic-ray background in practice.}

\begin{figure*}
    \centering
     \includegraphics[width=0.32\textwidth]{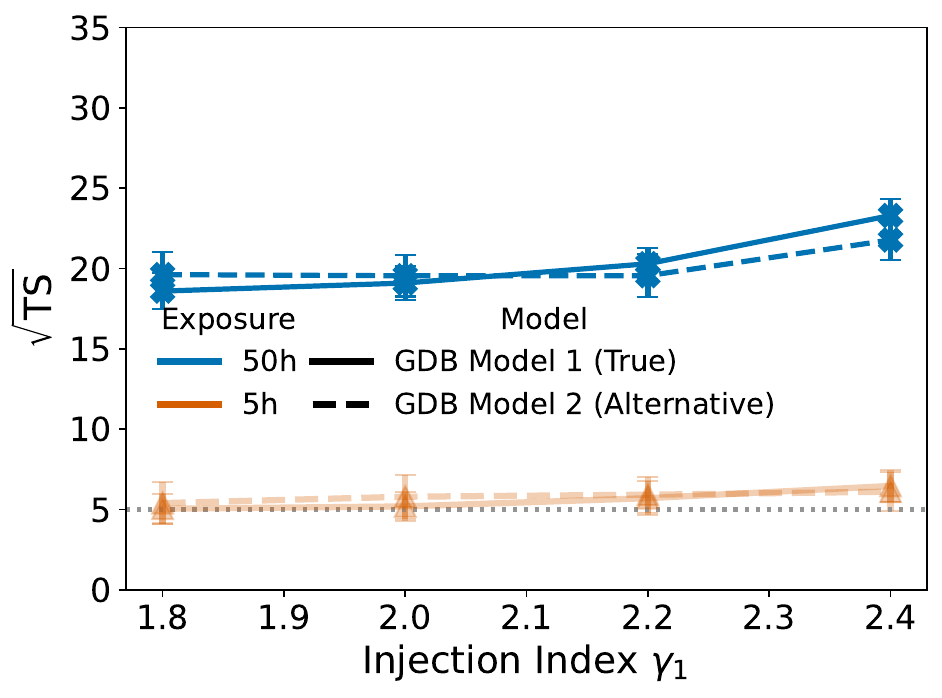}
    \includegraphics[width=0.32\textwidth]{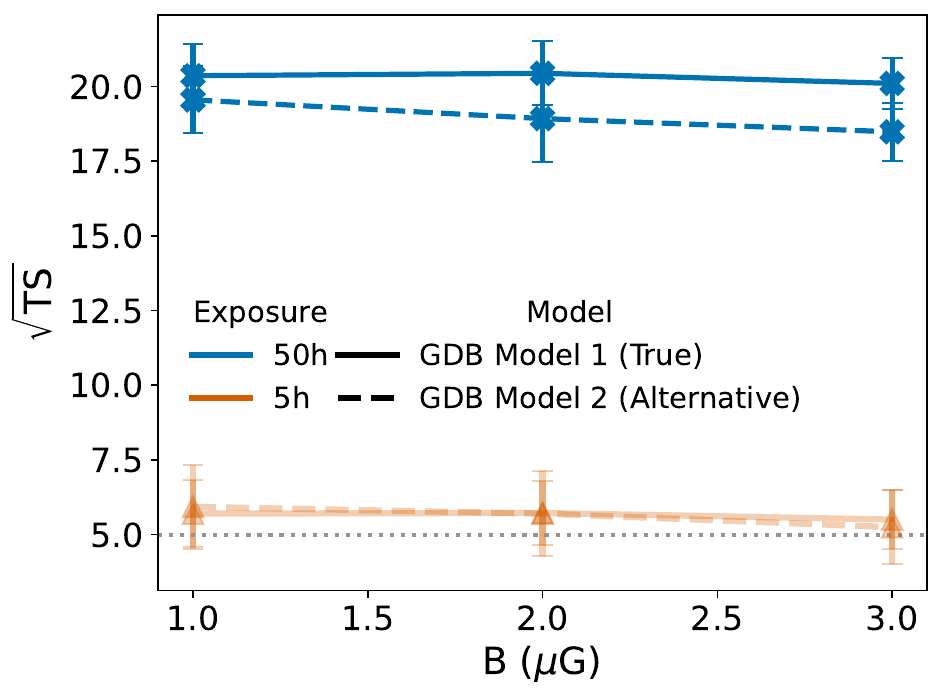}
    \includegraphics[width=0.32\textwidth]{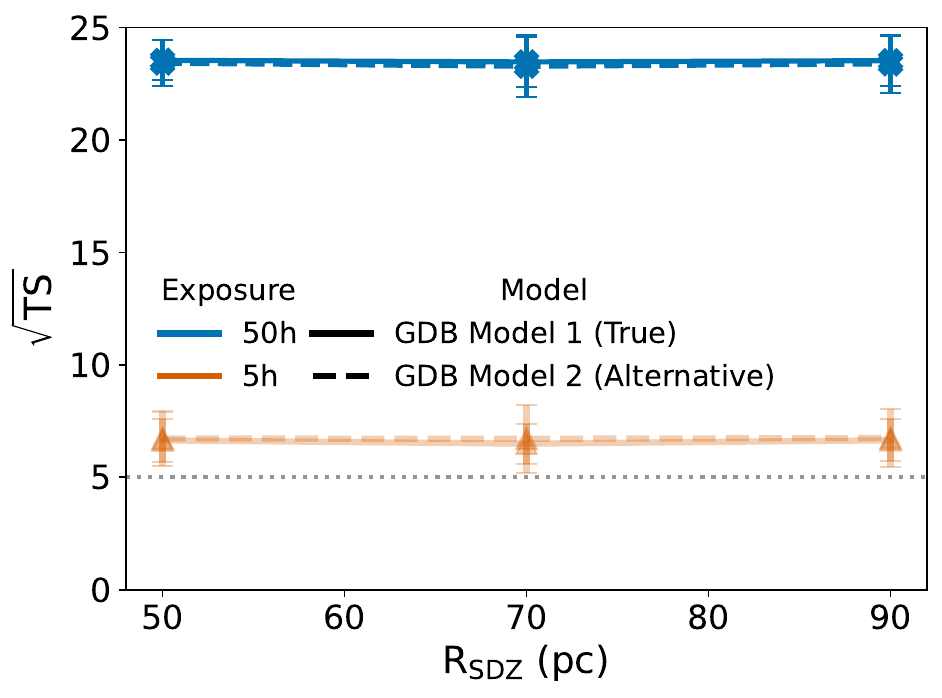}
    \caption{Detection significance of the Geminga halo by CTAO-South by fitting to the true astrophysical background and the alternative background. One model parameter is varied at a time while the remaining parameters are fixed at their benchmark values ($\gamma_1 = 2.2$, $B = 1\,\mu\mathrm{G}$, $R_{\rm SDZ} = 50$~pc). Results for a 5~h exposure are shown to illustrate how the detection significance improves with observing time; this represents an extremely conservative exposure for the CTAO. Each data point is derived from fitting 50 independent realizations of the same source model.  }
    \label{fig: Geminga alternative bkg South}
\end{figure*}

\begin{figure*}
    \centering
     \includegraphics[width=0.32\textwidth]{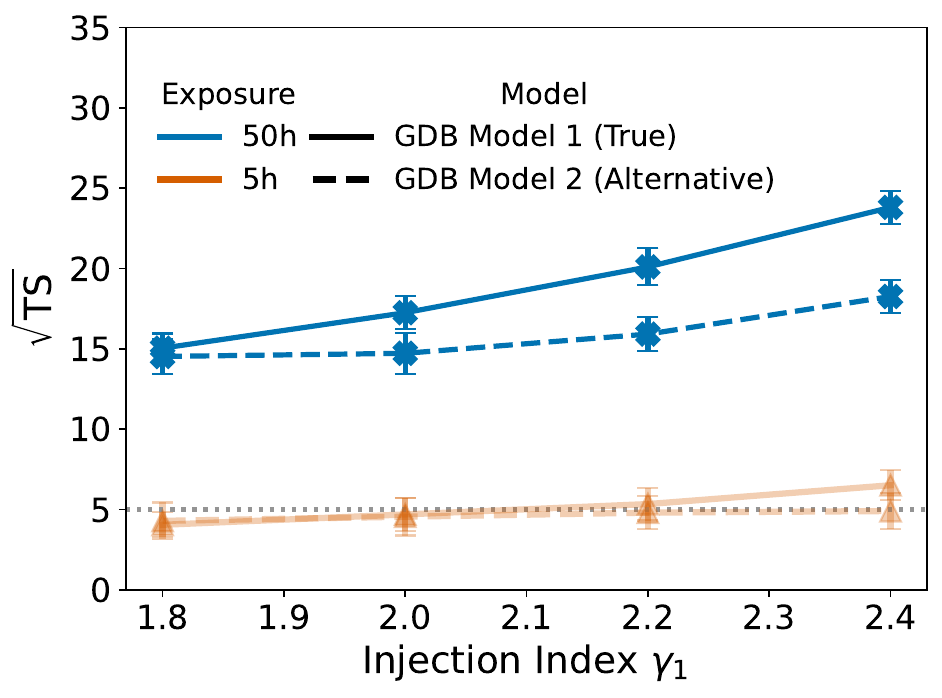}
    \includegraphics[width=0.32\textwidth]{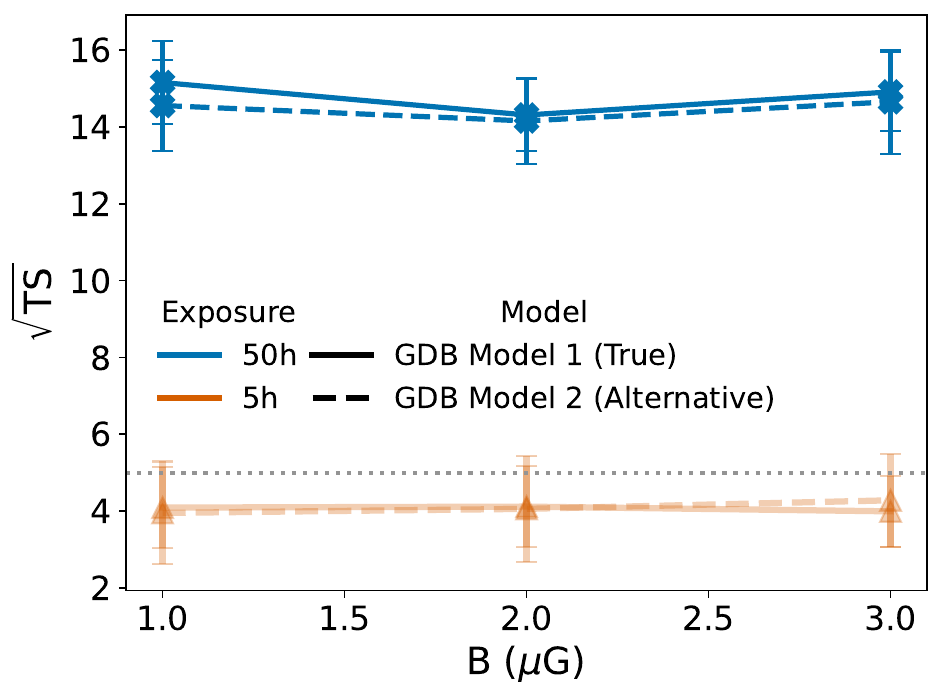}
    \includegraphics[width=0.32\textwidth]{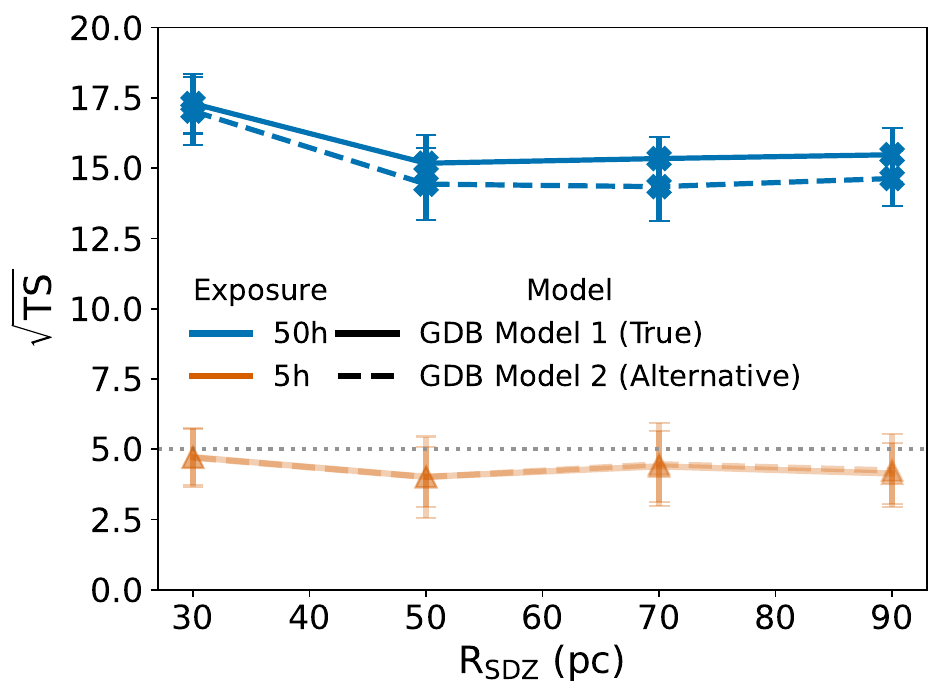}
    \caption{Detection significance of the Monogem halo by CTAO-South by fitting to the true astrophysical background and the alternative background. One model parameter is varied at a time while the remaining parameters are fixed at their benchmark values ($\gamma_1 = 1.8$, $B = 1\,\mu\mathrm{G}$, $R_{\rm SDZ} = 50$~pc). Results for a 5~h exposure are shown to illustrate how the detection significance improves with observing time; this represents an extremely conservative exposure for the CTAO. Each data point is derived from fitting 50 independent realizations of the same source model.  }
    \label{fig: Monogem alternative bkg South}
\end{figure*}

In this test, the mock observations produced with GDB Model~1 are analyzed using an alternate background prescription in which the diffuse component is replaced by GDB Model~2. The resulting detection significance, compared to the case where the ``true'' GDB Model~1 is used in the fit, is shown in Figure~\ref{fig: Geminga alternative bkg South} for Geminga, and in Figure~\ref{fig: Monogem alternative bkg South} for Monogem. Here, we refer to GDB Model~1 as the true background because it is the model used to generate the simulated observations. The results indicate that CTAO detections remain robust even when the diffuse background model is intentionally mismatched. For all tested source parameters, fitting the data with GDB Model~2 yields detection significances that closely track those obtained using the true GDB Model~1. Due to the limited FoV of the LST camera ($4.3^{\circ}$), which covers only the central region of the halo (see Figure~\ref{fig: intensity maps}), the statistics in the 20~GeV–150~GeV energy range are particularly sensitive to the diffuse background estimate. We therefore attribute the modest TS enhancement observed for $\gamma_1 = 1.8$ and $2.2$ in the Geminga case to a slight underestimation of the diffuse flux in the alternative background model, as indicated in Figure. \ref{fig:GDB compare}, which can artificially increase the apparent source contribution. The differences remain below $\sim 5\sigma$ for a 50~h observation, corresponding to $\lesssim 20\%$ variation in $\sqrt{\mathrm{TS}}$. Even for the shorter 5~h exposure, where the detection is near the threshold of $\sim 5\sigma$, the degradation from diffuse mismodeling remains modest.

\textbf{Irreducible cosmic-ray background}\\
Estimating the irreducible CR background remains a persistent challenge for IACT analyses of highly extended sources. In the H.E.S.S.\ detection of Geminga \citep{HESS_2023}, two complementary approaches were employed: an ON/OFF method using source-free observations, and a FoV background method in which the residual CR rate is modeled from Monte Carlo simulations of the instrument acceptance. \youyourev{For CTAO observations of TeV halos, this challenge persists across all telescope types. At SST energies ($\gtrsim 5$~TeV), where the halo emission contracts due to energy-loss-limited diffusion (see Figure~\ref{fig: intensity maps}), one might expect sufficient source-free area in the SST FoV for empirical background estimation. Defining a background region at $\sim$20--30\,pc from the pulsar would bias the integrated flux estimate by only $\sim$5--10\%, which would in principle still allow for a detection. The more fundamental obstacle is the mutual contamination between Geminga and Monogem: since both halos contribute flux across a large fraction of each other's FoV, a background region on the near side of either source would be contaminated by the neighbouring halo. Optimally, the background should therefore be sampled on the far side of the source relative to the neighbouring halo, for instance, using a half-ring geometry, to minimise this cross-contamination.} In the MST and LST energy ranges, the halo emission is even more extended and can fill a substantial fraction of the camera FoV. Consequently, the classical approach of defining source-free OFF regions within the FoV is not straightforward for either telescope type.
 
\youyou{In our simulations, mock observations are generated using the CTAO IRF-predicted irreducible background rate, implicitly assuming accurate knowledge of this component. In real observations, the accuracy of this prediction will need to be validated through independent means. A promising approach is cross-calibration with HAWC ($\sim$ 0.1--100\,TeV) and LHAASO ($\sim$ 0.1--1000\,TeV), which measure the source flux in the overlapping energy range with independent systematics. Since water Cherenkov detectors survey the full overhead sky and do not face the same FoV limitations, they provide robust external anchors for the source flux at SST energies. If the CTAO-measured source flux normalization in the overlapping energy range is consistent with the HAWC and LHAASO measurements, this provides confidence in the background calibration; a significant discrepancy would instead point to a mis-estimation of the residual CR background rate. At lower energies in the MST and LST range, where CTAO provides unique coverage below the HAWC threshold, no direct external anchor is available, and the background calibration relies on the IRF predictions together with the FoV normalization term included in our likelihood model.}
In practice, resolving the spatial morphology of the extended halo in the LST energy range may require dedicated observation strategies such as \youyourev{grid pointings}. By tiling multiple overlapping fields around the source, \youyourev{grid pointings} can map emission extending beyond a single camera FoV, which is necessary both to capture the imprint of pulsar proper motion on the low-energy morphology and to provide regions at the \youyourev{edges of the grid} where the source contribution is sufficiently low for empirical estimation of the irreducible CR background. \youyourev{This can be used to normalise the background components in the forward-folding approach applied in this study}.
\subsection{Sensitivity to transport parameters}
To assess the capability of the CTAO to constrain TeV-halo source parameters, we
generate mock observations of the Geminga and Monogem halos using the benchmark
model and simulate 5~h and 50~h pointed observations at CTAO-South. For each
configuration, we produce 50 statistically independent realizations to
characterize the impact of Poisson and background fluctuations.

We perform template fitting on each realization using both the benchmark model
(the “true” template) and alternative templates in which one parameter is
varied while the others are held fixed. The discriminability of the alternative
model is quantified using the likelihood ratio,
\begin{equation}
\Delta TS = 2 \left( \ln \mathcal{L}_{\text{true}} - \ln \mathcal{L}_{\text{alt}} \right),
\end{equation}
where $\mathcal{L}_{\text{true}}$ and $\mathcal{L}_{\text{alt}}$ are the
maximum likelihood values for the true and alternative templates,
respectively. We define
\begin{equation}
\sigma = \mathrm{sign}(\Delta TS) \cdot \sqrt{|\Delta TS|},
\end{equation}
such that positive $\sigma$ indicates preference for the true model and negative
values favor the alternative model. Although real analyses will face parameter
degeneracies and imperfect background knowledge, this forecast represents an
optimistic upper bound on CTAO constraining power.

\begin{figure*}
    \centering
    \includegraphics[width=0.32\textwidth]{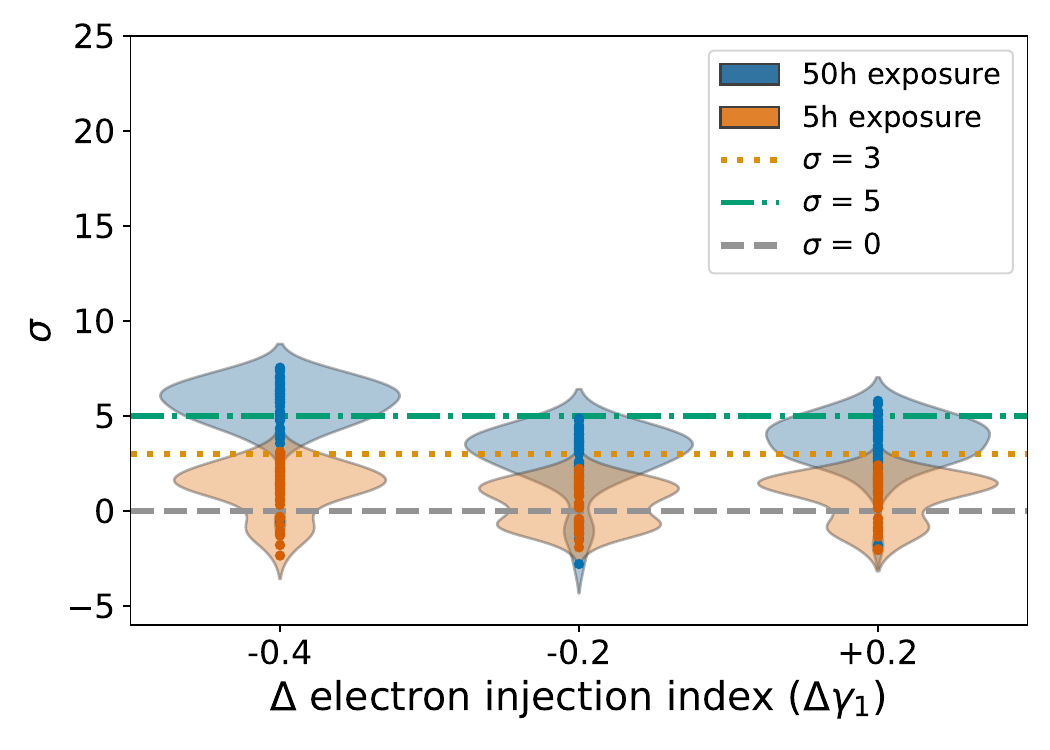}
    \includegraphics[width=0.32\textwidth]{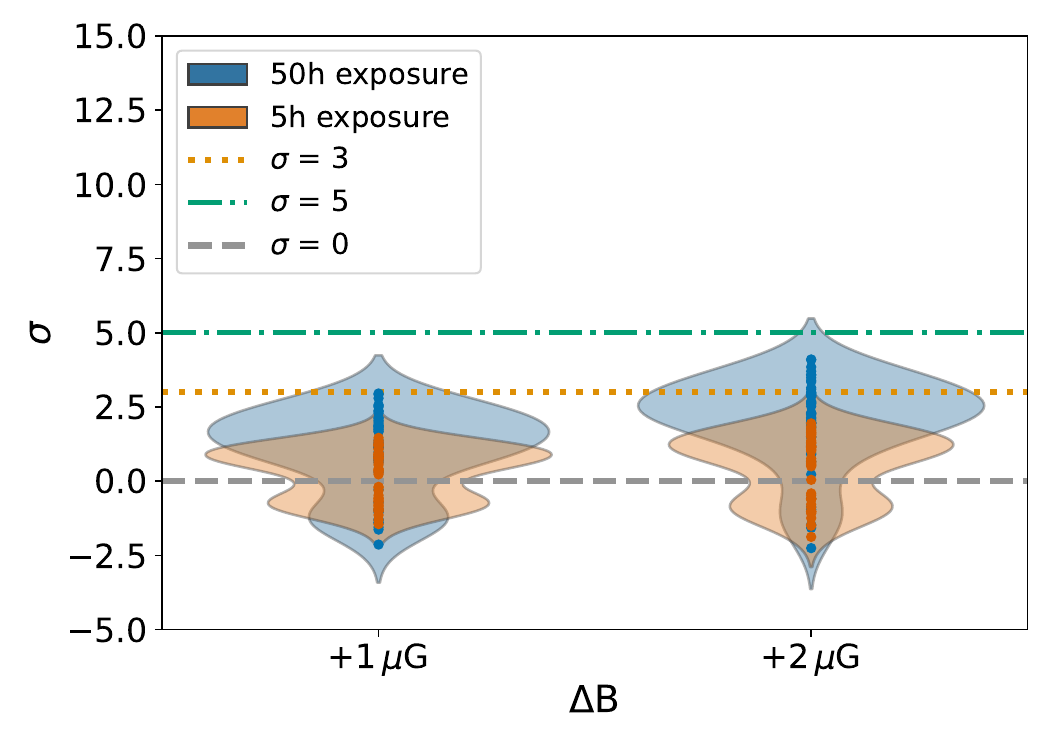}
    \includegraphics[width=0.32\textwidth]{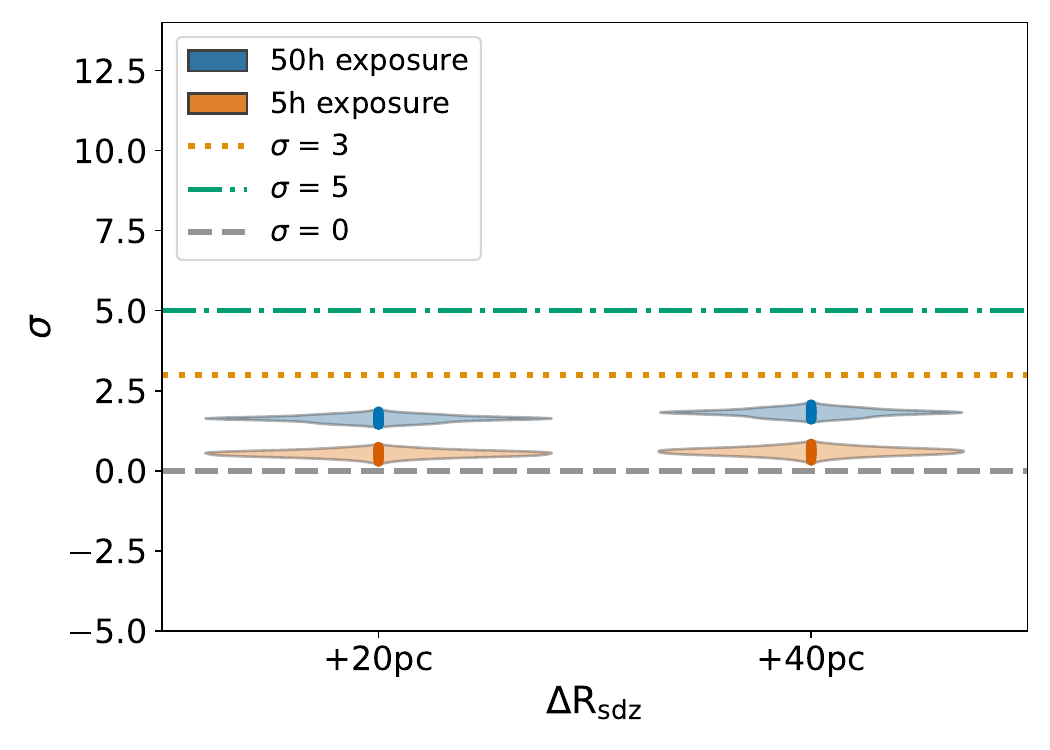}
    \caption{Discriminability of source parameters for the Geminga halo based on mock CTAO observations. Each panel shows violin distributions of the variability in the distinguishability metric $\sigma$ when varying a single source parameter: electron injection index $\gamma_1$ (left), magnetic-field strength $B$ (middle), and diffusion-zone radius $R_{\rm SDZ}$ (right). The true model (used to generate the mock data) is defined by $\gamma_1 = 2.2$, $B = 1~\mu\mathrm{G}$, and $R_{\rm SDZ} = 50$~pc. Blue violins correspond to 50~h exposure scenarios, while orange violins correspond to 5~h. Horizontal lines indicate reference levels of distinguishability at $\sigma = 0$ (dashed), $\sigma = 3$ (dotted), and $\sigma = 5$ (dash-dotted). Higher $\sigma$ values indicate greater statistical separation from the true model. 
    }
    \label{fig:Geminga_disdinguish}
\end{figure*}

\begin{figure*}
    \centering
    \includegraphics[width=0.32\textwidth]{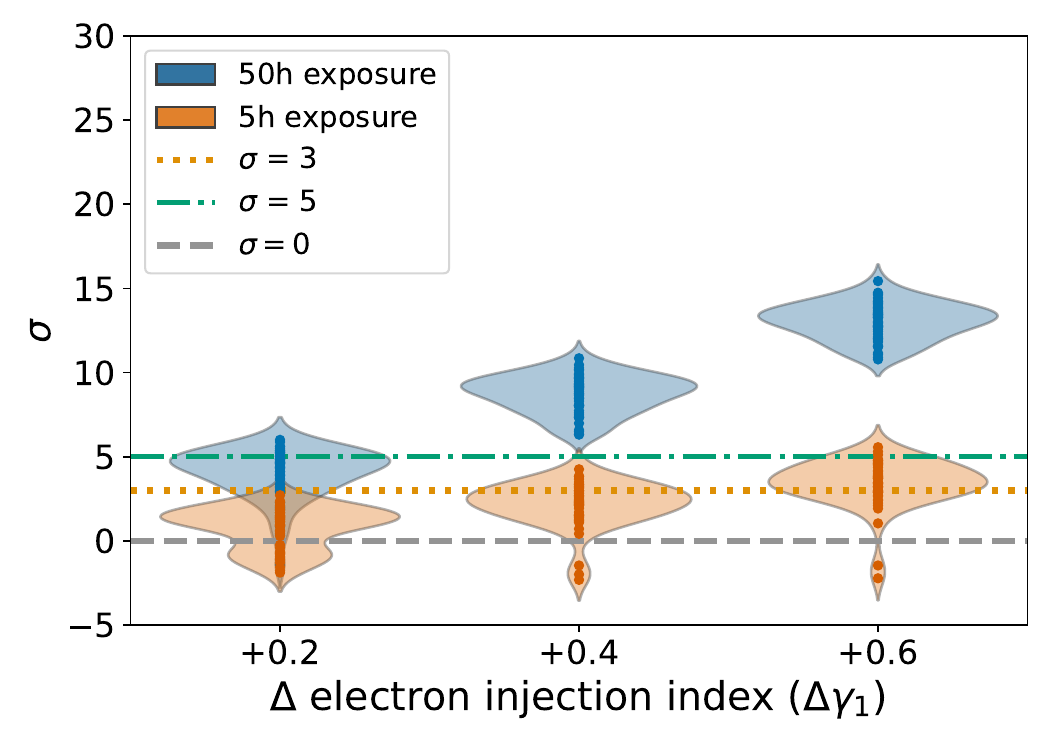}
    \includegraphics[width=0.32\textwidth]{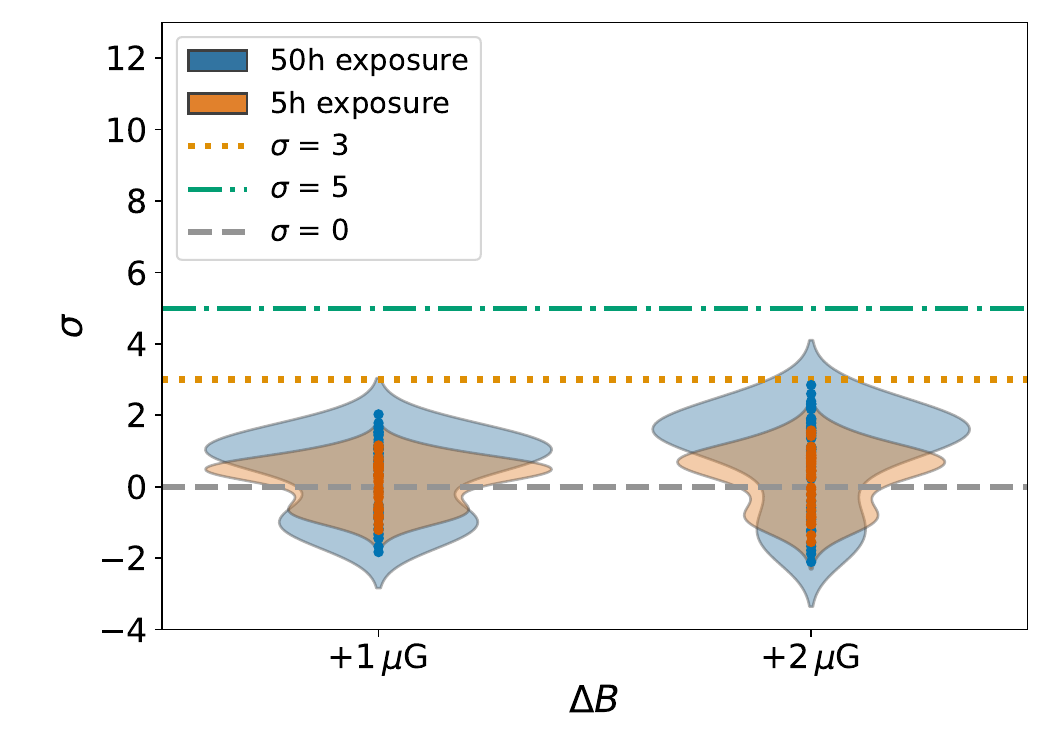}
    \includegraphics[width=0.32\textwidth]{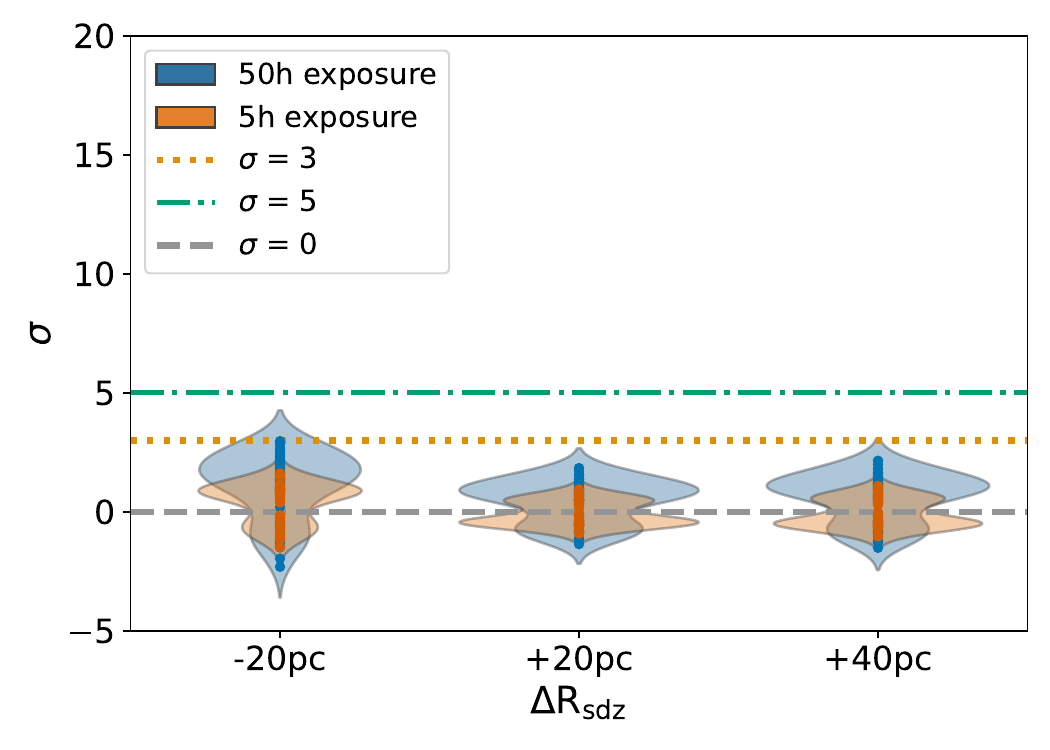}
    \caption{Discriminability of source parameters for the Monogem halo based on mock CTAO observations. Each panel shows violin distributions of the variability in the distinguishability metric $\sigma$ when varying a single source parameter: electron injection index $\gamma_1$ (left), magnetic-field strength $B$ (middle), and diffusion-zone radius $R_{\rm SDZ}$ (right). The true model (used to generate the mock data) is defined by $\gamma_1 = 1.8$, $B = 1~\mu\mathrm{G}$, and $R_{\rm SDZ} = 50$~pc. Blue violins correspond to 50~h exposure scenarios, while orange violins correspond to 5~h. Horizontal lines indicate reference levels of distinguishability at $\sigma = 0$ (dashed), $\sigma = 3$ (dotted), and $\sigma = 5$ (dash-dotted). Higher $\sigma$ values indicate greater statistical separation from the true model. }
    \label{fig:Monogem_disdinguish}
\end{figure*}

The resulting $\sigma$ distributions are shown in
Figure~\ref{fig:Geminga_disdinguish} (Geminga) and
Figure~\ref{fig:Monogem_disdinguish} (Monogem), where each violin summarizes the 50 realizations for a given parameter shift. With only 5~h exposure, the CTAO cannot
distinguish variations in injection index, magnetic field, or diffusion-zone size for either source. However, with 50~h exposure, the CTAO can discriminate
injection-index variations of $\Delta\gamma_1 \approx 0.2$ and magnetic-field variations at the level of $\Delta B \approx 2~\mu$G for both sources.

\youyou{The constraints presented above are obtained by varying one parameter at a time and therefore represent an optimistic upper bound on CTAO constraining power. We note, however, that these parameters are not intrinsically degenerate: each leaves a distinct imprint on the spatially resolved emission. The injection index ($\gamma_{1}$) primarily determines the spectral shape in the central region of the halo, the magnetic field ($B$) controls the radial gradient of the energy spectrum through the electron cooling rate, and the energy dependence of the emission extent is sensitive to the size of the slow-diffusion zone ($R_{\rm SDZ}$). Since our analysis employs spatially resolved morphological templates rather than integrated spectral energy distributions, these distinct spatial signatures are already \youyourev{tested} in the likelihood fit. The residual degeneracy arises from the use of a total test statistic summed over all energy bins, which dilutes the sensitivity of individual energy ranges where a given parameter has the strongest impact. In practice, performing the analysis in separate energy bands tailored to the regime where each parameter dominates could help reduce these degeneracies.}

\youyou{Constraints on the diffusion-zone size are subject to a different, geometric limitation.} For Monogem, 50~h observations allow discrimination between a compact SDZ
($R_{\rm SDZ} \approx 30$~pc), as suggested by HAWC, and a more extended case
($R_{\rm SDZ} \gtrsim 50$~pc). \youyou{However, once the spatial extent of the halo emission exceeds the CTAO FoV, the observations probe only the inner part of the halo. In this regime, different large values of $R_{\rm SDZ}$ produce very similar surface-brightness profiles within the mapped region, making them difficult to distinguish morphologically. Given the FoV sizes of the CTAO cameras (see Section \ref{subsubsec: CTAO config}), the SST, MST, and LST fields of view correspond to physical diameters of 44~pc, 40~pc, and 21~pc at the distance of Monogem. As the SDZ size increases, the halo broadens in the GeV–TeV range, which is primarily probed by the MSTs and the LSTs. For $R_{\rm SDZ} \gtrsim 50~\mathrm{pc}$, the resulting surface-brightness profiles vary only weakly within the CTAO FoV, rendering larger diffusion-zone radii effectively indistinguishable even with 50~h exposure.}
\section{Conclusions}
We performed end-to-end simulations of CTAO observations using two-zone transport models computed with \textsc{GALPROP}, together with realistic instrument response functions and diffuse-background uncertainties. Mock observations were generated and analyzed using a forward-folding likelihood framework to test both detection significance and parameter sensitivity for the Geminga and Monogem halos.

Our main findings are:
\begin{itemize}
    \item{\textbf{Robust Detectability:} Both Geminga and Monogem will be detected at high significance, reaching $\sim 15$–$30\sigma$ in 50~h pointed observations at CTAO-South and CTAO-North. Detection significance is stable against realistic diffuse-background mismodeling, with variations $<10\%$ in $\sqrt{\mathrm{TS}}$.}
    \item{\textbf{Spectral Constraints:} The electron injection index can be constrained to within $\Delta\gamma_1 \lesssim 0.2$, significantly improving current uncertainties.}
    \item{\textbf{Magnetic-Field Sensitivity:} the CTAO can distinguish magnetic-field strengths at the level of $\Delta B \sim 2~\mu$G.}
    \item{\textbf{Diffusion-Zone Morphology:} Compact diffusion zones ($R_{\rm SDZ}\approx 30$~pc) can be distinguished from more extended cases for Monogem, while diffusion radii $\gtrsim 50$~pc become increasingly degenerate due to the limited angular leverage within CTAO FoV.}
   \item  \youyou{\textbf{Background Calibration:} Our detection forecasts assume accurate knowledge of the irreducible CR background. In practice, the extended nature of both halos fills a substantial fraction of the CTAO field of view across all telescope types, precluding classical OFF-region background estimation within a single pointing. Cross-calibration with HAWC and LHAASO in the overlapping energy range offers a promising path to validate the background model at SST energies, while at lower energies, the background calibration will rely on the IRF predictions and the FoV normalization term in the likelihood fit.}
\end{itemize}
These results demonstrate that the CTAO will enable more precise measurements of the cosmic-ray transport properties in TeV halos.

\youyou{We emphasize that our sensitivity forecasts are conditional on the adopted two-zone diffusion framework. Within this scenario, we explored variations in key physical parameters, including the slow-diffusion-zone size, injection spectrum, and magnetic-field strength. This framework comes with a limitation to account for extra degrees of freedom, such as uncertainties in the proper motion, arbitrary orientations associated with anisotropic diffusion. In addition, properties of the magnetic turbulence itself—such as the coherence length—can influence the halo morphology through their impact on particle transport, representing a distinct physical effect from purely geometric anisotropies, as discussed by \citet{LopezCoto2018}. While a more comprehensive treatment of anisotropic transport and turbulence-scale effects would be a valuable extension of this work, the present study provides a physically motivated and internally consistent baseline estimate of CTAO sensitivity to TeV-halo transport parameters.}

Looking ahead, the CTAO will play a key role in probing particle injection and diffusion processes around pulsars. Realizing the full potential of these observations will require careful calibration of the irreducible cosmic-ray background and the modeling of the astrophysical backgrounds. Joint analyses with future radio and X-ray observations will help break degeneracies between magnetic-field strength and spatial transport, enabling a more complete understanding of the escaping-lepton population from PWNe. Extending this program to the growing sample of TeV halos discovered by LHAASO and HAWC will further clarify how common slow-diffusion environments are throughout the Galaxy.

\section*{Acknowledgements}
This work made use of Gammapy \citep{gammapy:2023}, a community-developed Python package, \youyourev{and the CTAO instrument response functions provided by the CTAO Consortium and Observatory (version prod5 v0.1; \citealt{CTAO_IRF_Prod5})}. OM acknowledges support from the U.S. National Science Foundation under Grant No. 2418730. 

\section*{Data Availability} 
The data underlying this article will be shared on reasonable request to the corresponding author.




\bibliographystyle{mnras}
\bibliography{mybib}


\appendix

\section{Electron Cooling Time}\label{appendix: cooling}
The cooling time of relativistic electrons due to synchrotron radiation in a magnetic field of energy density $U_{B}$ and inverse-Compton scattering in photon fields of energy density $U_{\rm ph}$ is given by: 
\begin{equation}\label{eq:cooling_time}
    \tau_{\rm cool}=\frac{3m_{\rm{e}}c^{2}}{4c\sigma_{T}\gamma}\left(U_{B}+
            \sum_i \frac{U_{\text{ph}, i}}{(1+4\gamma\varepsilon_{0, i})^{3/2}}\right)^{-1},
\end{equation}
where $m_{\rm e}$ is the electron mass, $c$ is the speed of light, $\sigma_{\rm T}$ is the Thomson scattering cross section, $\gamma$ is the electron Lorentz factor, and $\varepsilon_{0,i}$ is the ratio of the mean photon energy of the $i$th radiation field to the electron rest-mass energy. The Klein--Nishina (KN) effect becomes significant when
$\gamma \gtrsim \gamma_{\rm KN} \equiv (4\varepsilon_{0})^{-1}$, leading to a suppression of the IC cooling rate at high energies. 

In our calculations, we include the following components of the interstellar radiation field: optical ($U_{\rm opt}=0.56$~eV~cm$^{-3}$, T$_{\rm opt}=5000$~K), infrared ($U_{\rm IR}=0.41$~eV~cm$^{-3}$, T$_{\rm IR}=20$~K), and the Cosmic Microwave Background ($U_{\rm CMB}=0.26$~eV~cm$^{-3}$, T$_{\rm CMB}=2.7$~K) \citep{john2023}. The cooling time of a 100~TeV electron in a $\{1,2,3\}~\mu$G magnetic field is then $\{22.8, 14.8, 9.3\}$~kyr.

\section{Source Zenith Angles}\label{appendix: zenith}
To assess the observational conditions of the Geminga and Monogem pulsars from both CTAO sites, we calculate their zenith angles as a function of time throughout a year. The zenith angle, defined as the angle between the source direction and the local zenith at the observatory, directly impacts the instrumental response of the CTAO, influencing the effective area, angular resolution, and energy threshold of the
observations.

\begin{figure}
    \centering
    \begin{subfigure}[b]{0.5\textwidth}
        \centering
        \includegraphics[height=0.20\textheight]{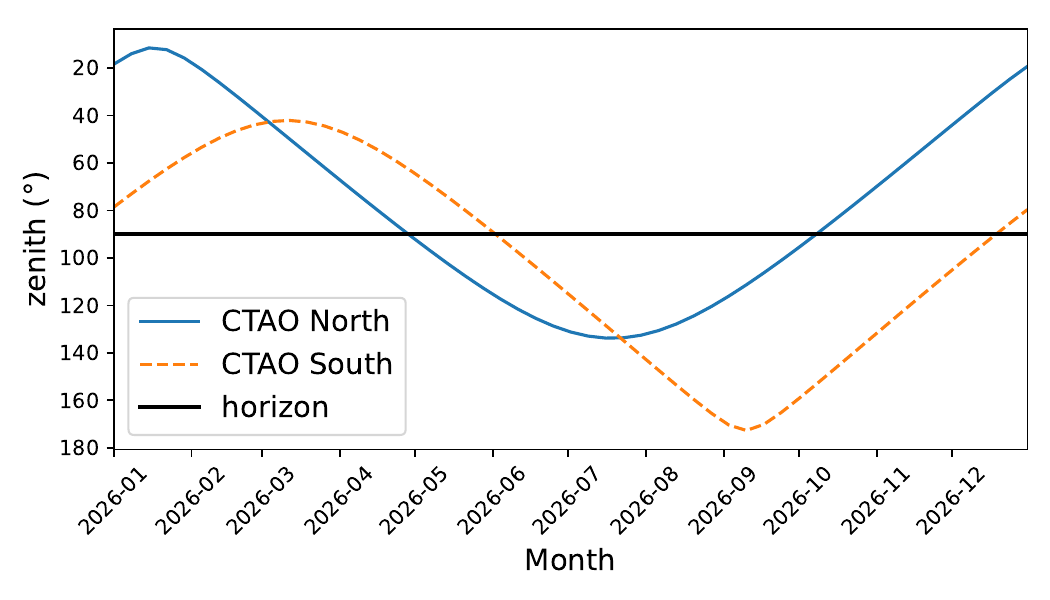}
     \end{subfigure}
     \hfill
     \begin{subfigure}[b]{0.5\textwidth}
         \centering
         \includegraphics[height=0.20\textheight]{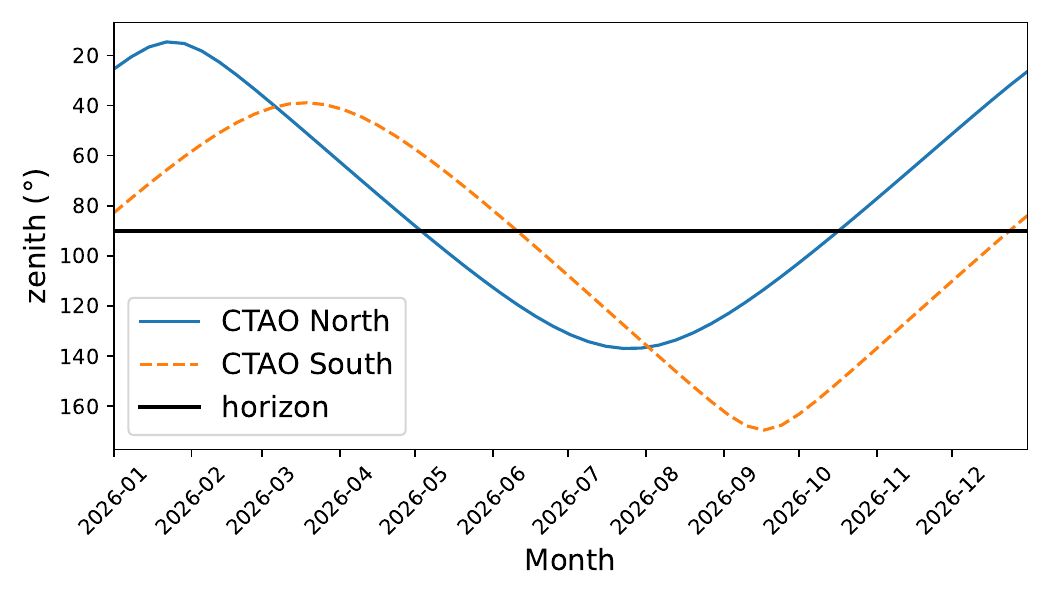}       
     \end{subfigure}
\caption{Zenith angle of Geminga halo (\textbf{top panel}), and Monogem halo (\textbf{bottom panel}) from CTAO-North and South.}
     \label{fig: zenith}
\end{figure}

Figure~\ref{fig: zenith} shows the zenith-angle evolution of the
Geminga halo (top panel) and the Monogem halo (bottom panel) as seen from CTAO-North and CTAO-South.

\bsp	
\label{lastpage}
\end{document}